\def\half {{1\over 2}}
\def\ref  {}

\def\ket#1{\vert #1 \rangle}
\def\braket#1#2{\langle#1 \vert #2\rangle}

\documentstyle[preprint,epsf,aps,version2]{revtex}

\begin{document}

\begin{title}
Time, Gravity, and Quantum Mechanics
 \end{title}

\author{W. G. Unruh}
\begin{instit}
 CIAR Cosmology Program\\
Dept. of Physics\\
University of B. C.\\
Vancouver, Canada V6T 2A6\\
\end{instit}

\begin{abstract}

The role that time plays in Einstein's theory of gravity and in quantum
mechanics is described, and the difficulties that these conflicting roles
present for a quantum theory of gravity are described.
\end{abstract}

\section{ Gravity and Time}

The relation of any fundamental  theory to time is  crucial as was evident
from the earliest days of physics. If we go back to Newton's
{\it Principia}, in which he established   a general theoretical structure
 for the field of physics,   we find an odd series of sentences in the
first
few pages. He tells us that it is unnecessary  to define time as
it is obvious to all, but then proceeds to do just that. And his definition
is, certainly to modern eyes, rather strange. To quote\cite{1}

\begin{quote}
... I do not define time, space, place, and motion as being well
known  to all.  Only I observe, that the common people conceive these
quantities
under no other notions but from the relation they bear to sensible objects.
And thence arise certain prejudices, for the removing of which it will
be convenient to distinguish them into absolute and relative, true and
apparent, mathematical and common.

{I. Absolute, true and mathematical time, of itself, and from its
own nature, flows equably without relation to anything external, and by
another name is called duration: relative, apparent, and common time, is
some sensible and external (whether accurate or unequable) measure of
duration by means of motion, which is commonly used instead of true time;
such as an hour, a day, a month, a year.}
\end{quote}

Reading this definition today, it is hard to see what the fuss is about.
Time
for
us common folk is exactly Newton's true time. Taught about
time since we were small, we know that there is something, insensible
but present in the universe, called time, something that is separate from
the other objects in the universe and the same everywhere. Newton would
certainly not need to include his definition today, nor would he ascribe
to common man that which he did.

It is precisely
because we have so thoroughly absorbed Newton's lesson that we all have
immense difficulty in coming to terms with the revolutions in physics
 of the twentieth
century and that we in physics now have such difficulty in producing
a unified theory of quantum mechanics and gravity.

It is the purpose of this paper to sketch the changes in the notion of
time
in twentieth century physics. I will show how in General Relativity the
nature of time changed radically that of Newton and how
gravity itself became an aspect of time. I will then examine the role time
 plays  in quantum physics. Finally, I will
sketch ways in which the lack of a  proper understanding of time
seems to be one of
the chief impediments to developing a quantum theory of gravity.

The change  began with Special Relativity, the first theory
 in which time lost some part of its absolute and invariant character.
 Time became, at least in some small
sense, mutable. It was precisely this conflict between a mutable notion
of time and the  absolute and unitary notion of time inherited from Newton
 that has caused consternation and confusion. This
confusion came about not because of any innate violation of the sense of
time that we are born with. Time for children is flexible and changeable,
 and certainly need not be the same here as it is there. Throughout
our early years we were taught
 the lessons of Newton. Time was something out there,
something that our watches measured, and something that really was the
same everywhere. We learnt while very young that our excuse to the teacher
that our time was different from the teacher's
time was not acceptable.

 The conclusions of Special Relativity
came into direct conflict with these early lessons. The `Twins Paradox'
  is  the epitome of this confusion, because there is, of course,
no paradox at all except in the conflict between the notion of time as
expressed in this theory and the notion of time as expressed by Newton.
It is because we have so thoroughly absorbed Newton's definition of time
 that we become
confused when time in Special Relativity behaves differently.
In Special Relativity time, at least time as measured by any physical
process, became not the measure of that unitary non-physical phenomenon
of  universal time, but a measure of distance within the new construct
of `space-time'. No-one expresses any surprise, or considers it
a paradox, that the difference  in the odometer
readings on my car from the start of a trip to its end is  not the same
as the
difference on your odometer for  our trips from Vancouver to
Toronto, especially if I went via Texas, while you went via
the Trans Canada
Highway. Distances are simply not functions only of the end points of the
 journey but depend on
 the complete history of the journey. Within Special Relativity the
same is true of
time. Times are no longer  dependent only on the beginning and end of our
journey,
but are  history-dependent and depend on the details of the
 journey themselves in exactly the same
way that distances do. Mathematically this is expressed by having the
time in Special Relativity be given
by an extended notion of distance in an extended space called space--time.
Just as
 the spatial distance between two points depends on the details of the
path
joining the to points, so the temporal distance joining two
 points at two separate
instants depends on the details of the path and the speed along that path
joining the two points at the two instants.

Even though time as a measure of the duration of a process  became  mutable,
Special Relativity
was still a theory which retained some of the Newtonian picture. Just
as space, for Newton, was another of those non-material but real invariant
externals, so in Special Relativity space-time is also a real non-material
 invariant external. Instead of having two such concepts, i.e.,
space and time, Special Relativity has them  unified into one single concept
 which retains most of the features of space.

This changed  in General Relativity, Einstein's theory of gravity.
Within Special Relativity, the immutability, the sameness, the independence
 of space and time  from the other attributes
of the universe, was kept inviolate. Although time, as measured by a watch,
was path-dependent, it was the same everywhere, and was independent of
the
nature of matter around it. In General Relativity this aloofness
vanished.

One often hears that what General Relativity did was to make time depend
on gravity. A gravitational field
{\it causes} time to run differently from
one place to the next, the so called `gravitational red shift'. We hear
about
experiments like the one done by Pound and Rebka \cite{3}
in which the oscillation frequency of Iron Nuclei at the top and the
bottom of the
Harvard tower were found to differ, or
about Vessot's\cite{4} `disposal' procedure
of one of his Hydrogen masers in which such a
 maser was shot up 10,000 km above
the earth  by rocket
before dropping into the Atlantic. During that journey, he noted that
the maser
ran more quickly at the top of its trajectory than at the
bottom, in agreement with General Relativity to better than one part in
five thousand. The lesson of these experiments would appear to be that
gravity
alters the way clocks run.  Such a  dependence  of time
 on gravity would have been strange enough
for the Newtonian view, but General
Relativity is actually much more radical than that. A more accurate way
of
 summarizing the lessons of  General Relativity is that
gravity does not {\it cause} time to run differently in different places
(e.g., faster far from the earth than near it). Gravity
{\it is} the unequable flow of time from place to place. It is not that
there
are two separate phenomena, namely gravity and time and that the one, gravity,
 affects
the other. Rather the theory states that the phenomena we usually
ascribe to gravity are actually caused by time's flowing unequably from
place to place.

This is strange. Most people find it very difficult
 even to imagine how such a statement could be true. The two concepts,
time
and gravity, are so different that there would seem to be no way that they
could possibly have anything to do with each other, never mind being identical.
That gravity could affect time, or rather could affect the rate at which
clocks run, is acceptable, but that gravity is in any sense the same as
 time seems naively unimaginable.
To give a hint about how General Relativity accomplishes this identification,
I will use  an analogy. As with any analogy, there will
be certain features that will carry the message that I want to convey,
and I will emphasize these. There are other features of the analogy which
may
 be misleading, and I will point out a few of these. The temptation with
any  analogy is to try to extend it, to think about the subject (in
this case time and gravity) by means of the analogy and to ascribe to
the theory (General Relativity) all aspects of the analogy, when in fact
only
some of the aspects are valid.

In this analogy I will use the idea from Special Relativity that some of
the
 aspects of time
are unified with those of space and that the true structure of space-time
is in many   ways the same as our usual notions of space. I will therefore
use a spatial analogy to examine certain features of space-time in the
vicinity
 of the earth.
In order to be able to create a visual model, I will  neglect two
of the ordinary dimensions of space and will be concerned only with the
physical spatial dimension of up and down along a single line through the
center of
  the earth chauvenistically chosen to go through my home city of Vancouver.
In this model, time will be
represented by an additional spatial dimension, so that my full space--time
model will be given by a two dimensional spatial surface. What   I will
now
argue is that
I can explain the most common manifestation of gravity that when
I throw something up into the air, it slows down and finally stops its
ascent
and then  comes back
down to the surface of the earth (i.e., that which goes up must come down).
Usually one ascribes this behaviour to the presence of a force called gravity
which acts on the object, pulling it down toward the center of the earth.
The crucial point is that  one can alternatively explain this
 essential attribute
of gravity by assuming that time flows unequably from place to place,
without calling into play any `force of gravity' at all.

\epsfysize=3in
\centerline{\epsfbox[100 250 576 600]{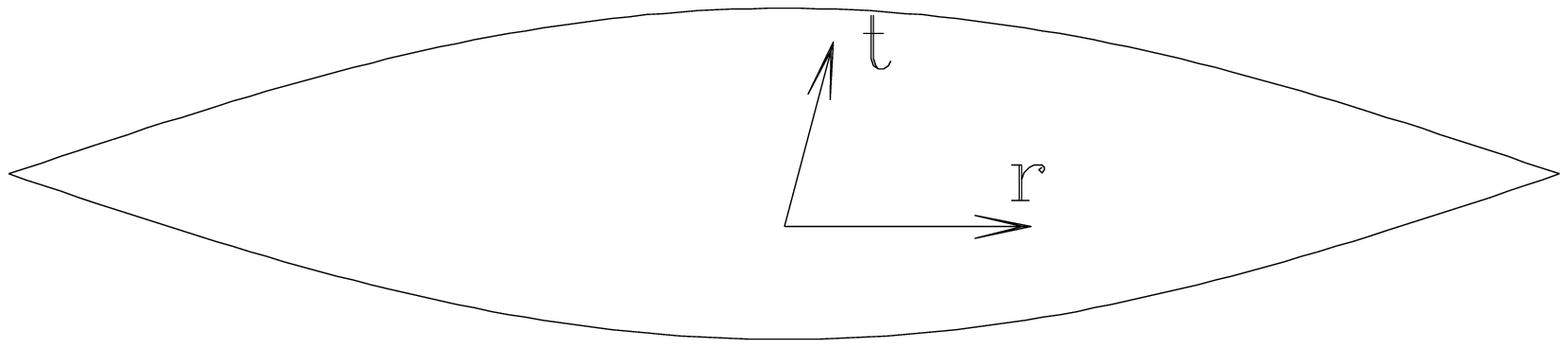}}
\centerline{Figure 1}
\centerline{\it The lens shape used to create the model of the
space-time near the earth}
\bigskip

In order to develop the analogy, we must first interpret the phrase
`time flows unequably'
 in terms of our model. We can assume, for the purposes of  our discussion,
that the physical phenomena near the earth are the same from one time to
the next. I.e., if we are to construct the two-dimensional space-time
 (up-down,
and time) near the earth out of pieces representing the space-time at
different instants, those pieces should all look the
same. I will use pieces that look like lenses (see figure 1).
The direction across the lens I will take as time, and the direction along
the lens I will take as space. These lenses have the feature that the
physically measurable time, which  Special Relativity teaches is just the
 distance across
the lens, varies from place to place along the lens.
I have thus interpreted the phrase
 `the flow of time is unequable'
as the statement that the distance (time) across each
lens is unequal from place
to place. I will
now glue a large number of these lenses together along their long sides,
giving us the  two dimensional shape in figure 2, for which
I will use the technical
 term `beach
ball'.  I will take this beach ball to be a  model of
 the space-time near the earth,
 a space-time made out of pieces on which time flows
 unequably from place to place.

\epsfysize=3in
\centerline{ \epsfbox[50 100 624 676]{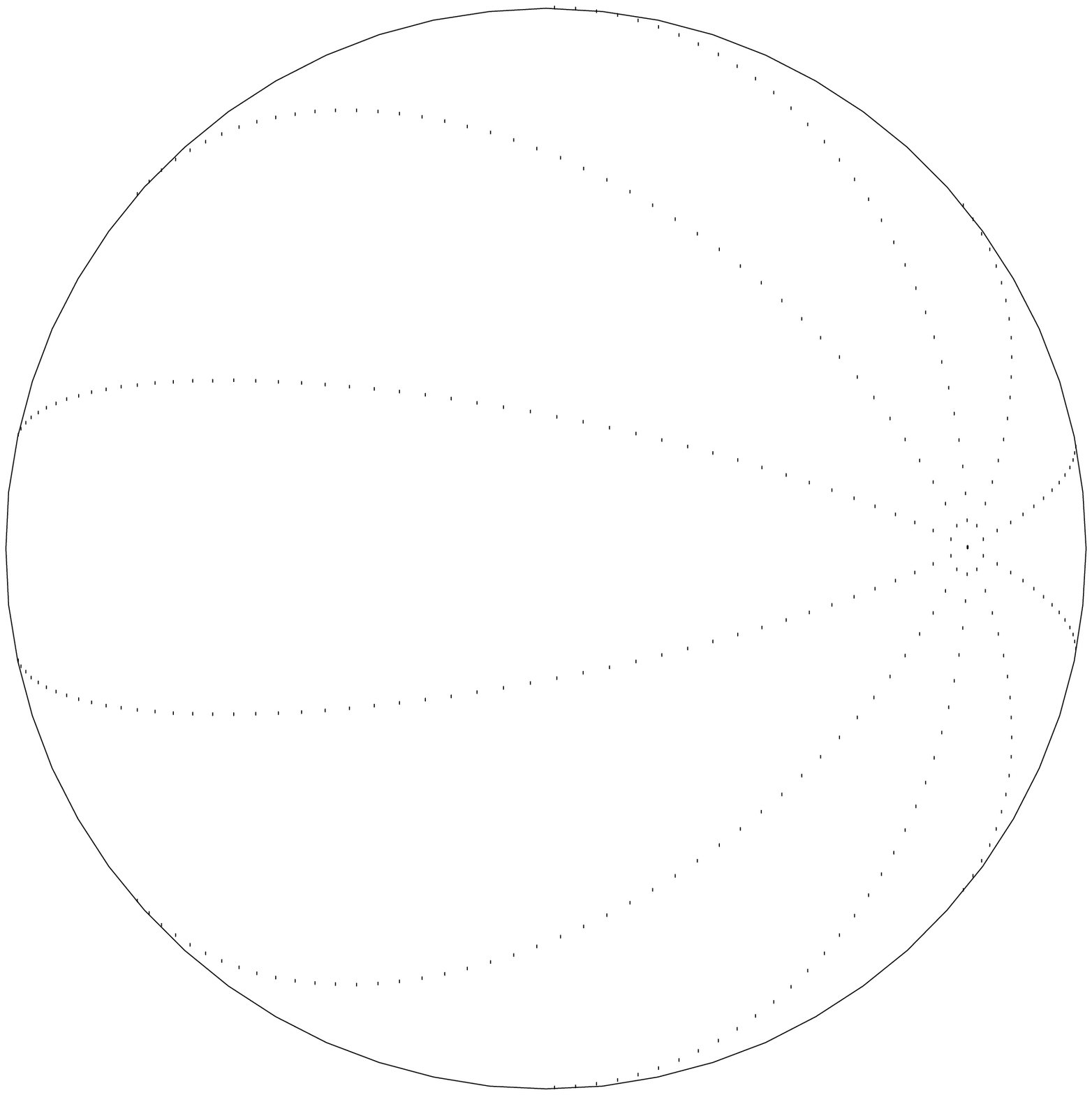}}
\centerline{Figure 2}
\centerline{ \it The spacetime made by gluing together the lenses}

\bigskip
Let us now locate the surfaces of the earth in this model. The earth
in the up--down direction has two surfaces-- one here in Vancouver
and the other one near  Isle Crozet in the south Indian Ocean. Since
the
distance through the earth from Vancouver to this island is constant, the
distance between the two strips on the surface of the beach ball must
be constant from one time to the next to model this known fact about
the earth  accurately.  Furthermore, since we expect the
system
to be symmetric, we expect that `up' here at Vancouver, and `up' at Isle
Crozet
should behave in exactly the same way. Thus,
the strips should be placed symmetrically on the beach ball. I thus have
figure 3, with the two black strips representing the two surfaces of the
earth
and the region between the strips representing the interior of the earth.

\epsfysize=3in
\centerline{\epsfbox[50 100 624 676]{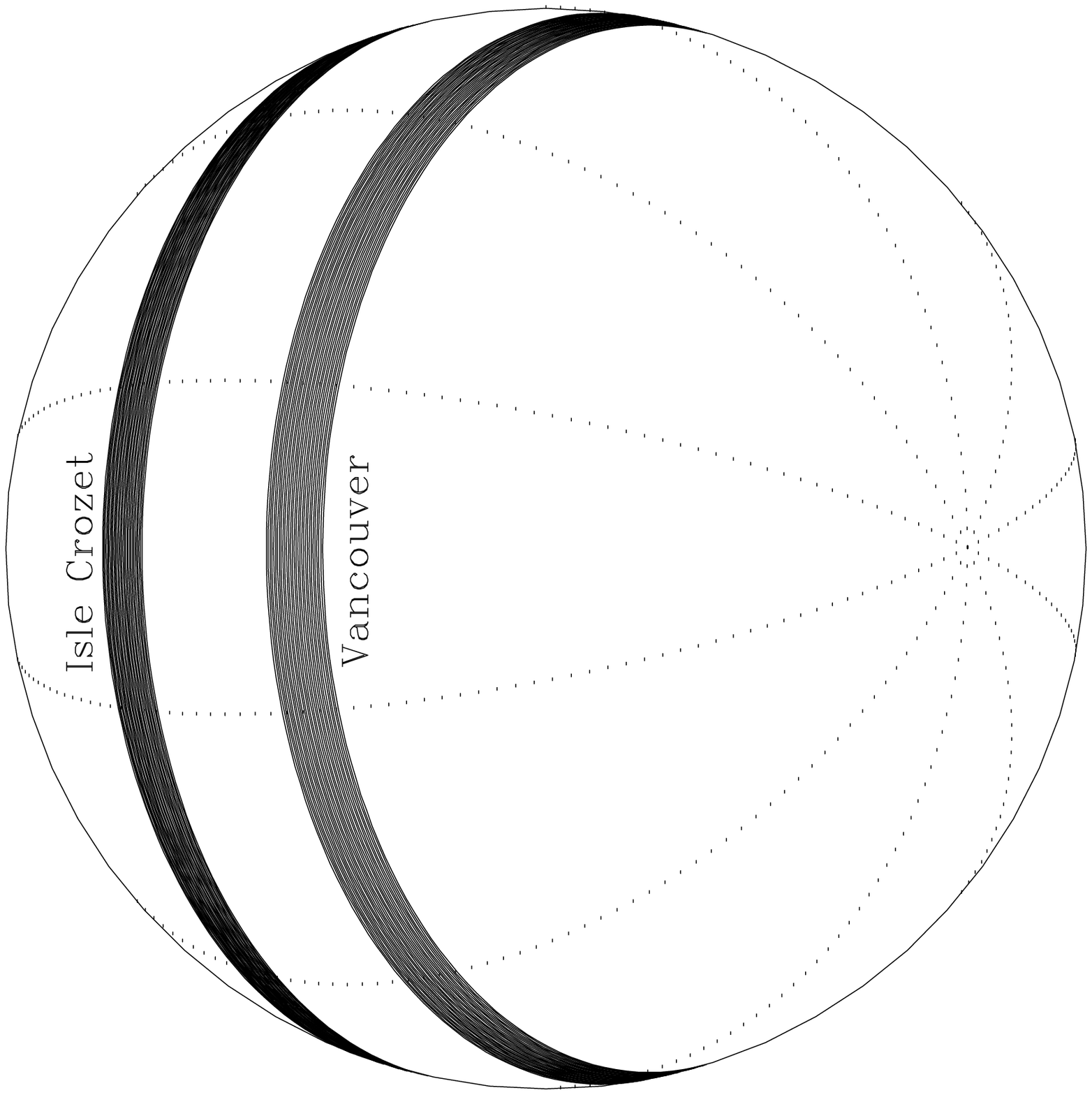}}
\centerline{ Figure 3}
\centerline{\it The two surfaces  of the earth  at
 Vancouver and at Isle Crozet drawn }
\centerline{\it in on the spacetime model. Between the two
strips is the interior of the earth.}
\bigskip

I stated that I would use this model to explain why, when something is
thrown into the air, it returns  to the earth. To do so, we must first
decide how bodies move when plotted in this spacetime. I go back to the
laws
of motion first stated by Newton, in particular his first law of motion.
\begin{quote}
{Every body continues in its state of rest, or of uniform motion
in a right line, unless it is compelled to change that state by forces
impressed upon it.}
\end{quote}

As I stated above, I will dispense with the idea of a gravitational force.
 I want to describe gravity  not as a force but as  the
unequable flow of time from place to place. A body
 thrown up into the air is
thus not acted upon by any forces. By the above law the body will
 continue its motion in
a `right',
or as we would now say, a straight line. But what does `straight'  mean
in this context of plotting the path of the body on the surface of
 this beach ball?  I  go back
to one of the oldest definitions of a straight line, namely that a  line
is straight
if it is the shortest distance between any two points along the line.

The beach ball is the surface of a sphere. On the surface of a sphere the
shortest distance between two points is
given by a great circle. Thus, applying this generalization of Newton's
Law,
 the free motion of a body on the two dimensional
space-time modeled by the beach ball will be some great circle around the
ball. If we plot the vertical motion of an object thrown into the
 air at Vancouver on
this model of the space--time, that plot will have the object
 following a great circle (straight line) on the surface of the
 beach ball. Consider the
 great circle given in figure 4. Starting at point A,
it describes the behaviour of  a particle leaving the surface of the earth.
 Initially, as time increases, the particle goes progressively further
 from the surface of the earth. As gravity is not a force, the particle
 continues to travel along the straight
line (great circle).  Eventually, the distance from the earth stops
increasing and begins to decrease.  The straight line,  the great circle,
 re-intersects the band representing
the surface of the earth at Vancouver  at point B.  I.e., the particle
has
returned to
the surface of the earth, just as a real body thrown away from the surface
of the earth will eventually return thereunto.

\epsfysize=3in
\centerline{\epsfbox[50 100 624 676]{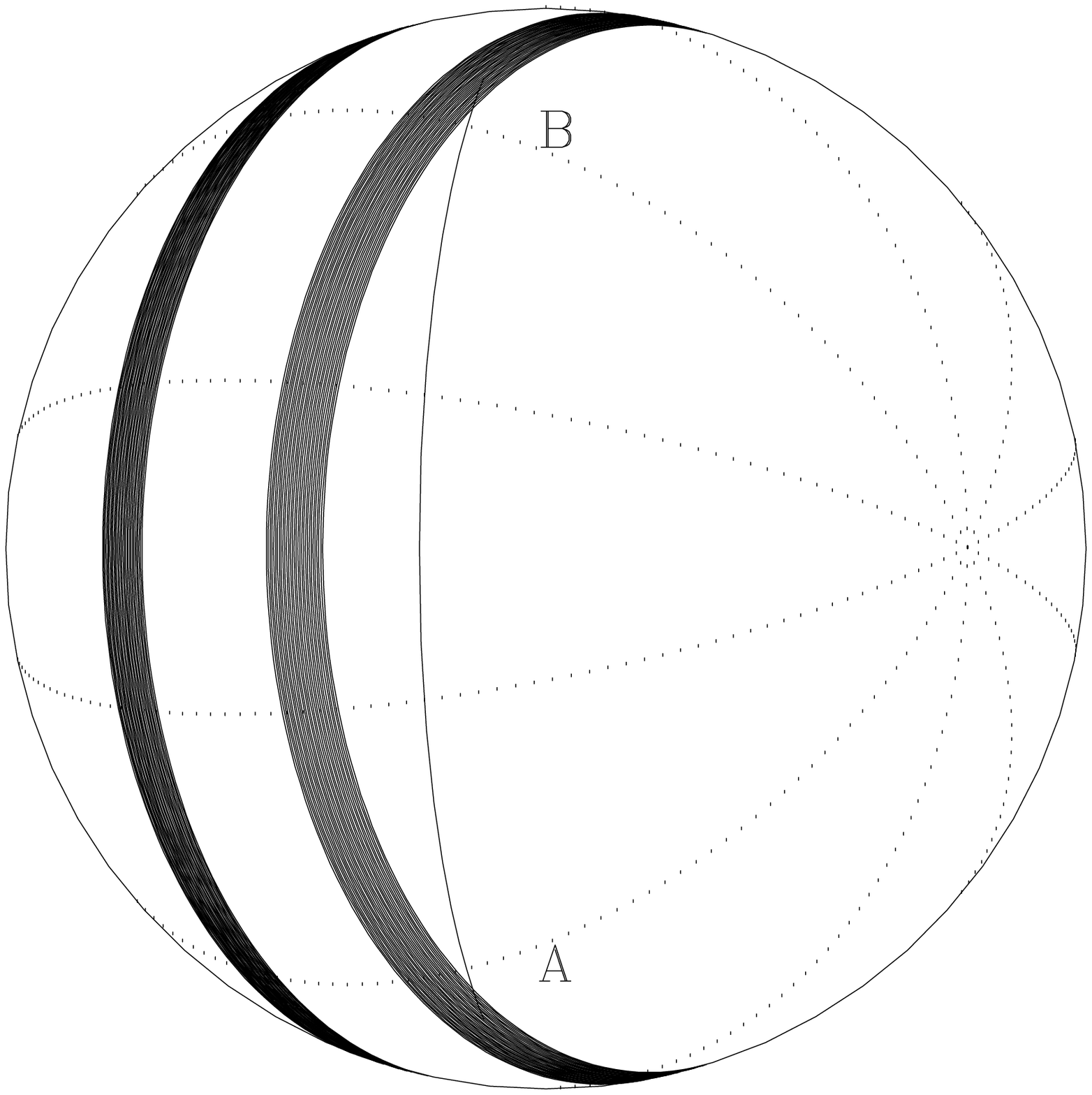}}
\centerline{Figure 4}
\centerline{ \it A great circle leaving the earths surface at {\rm A} and
returning at {\rm B}.}
\bigskip

Although one cannot construct as simple a model of the situation, one can
show that the same concept of the unequable flow of time can describe the
 behaviour of the moon  as it orbits the earth. The moon also is following
 a straight line through the spacetime surrounding the earth, a spacetime
constructed so that the flow of time far from the earth differs from its
flow near the earth. The line is certainly not straight in space, but it
is straight if plotted in space--time, straight in the sense of always
following a path which either maximizes or minimizes the
distance between any two points
 along the path.

With  the above  simple two-dimensional model one  can  also explain another
aspect of gravity,
namely the pressure
we feel on the soles of our feet as we stand.
Usually we ascribe this pressure to the response of the earth to the force
of
gravity pulling us down. As Einstein
already pointed out in 1908, there is another situation in which we feel
the same pressure, namely in an elevator accelerating upwards. In that
case the
pressure is not due to the resistance of the floor of the elevator to some
force pulling us down; rather, it is the force exerted
on us by the elevator in accelerating us upwards.
 Thus another possible
explanation for the force we feel under our feet is that the surface
of the earth is accelerating
upwards. Of course the immediate reaction is that this seems silly--- if
the
earth at Vancouver were accelerating upwards and that at Isle Crozet were
also accelerating upwards (since people there also feel the
 same force on the soles of
their feet when they stand), the earth must surely be getting larger. The
distance between two
objects accelerating away from each other must surely be changing. In the
presence of an unequable flow of time this conclusion does not necessarily
follow, as I can again demonstrate with the beach ball. Both sides of the
earth can be accelerating upwards even though the distance between them
does not change.\cite{curved}

In our beach ball model, the diameter of the earth (the distance
 between the two black lines) is clearly constant
 at all times. Let me carefully cut out one of the black
strips, the one representing the surface of the earth at Vancouver
 say, as in figure
5.  I will lay the strip out flat, as I have done with the peeled
portion of the strip in figure 5. The resulting graph is
just the same as the graph of an accelerating object in flat space-time.
Local (within the vicinity of the strip)  to the surface of the earth,
the space-time is the same as that around an accelerating
particle, and one can therefore
state that the surface of the earth at Vancouver is
accelerating upwards. It is not
following a straight line, It is following a curved line,
and by Newton's first
law must therefore have a force ( the one we feel on the
soles of our feet) which
causes that acceleration. It is
the acceleration upward of the surface of the earth which leads to the
sensation of a force on the soles of our feet.

\epsfysize=3in
\centerline{\epsfbox[150 200 524 600]{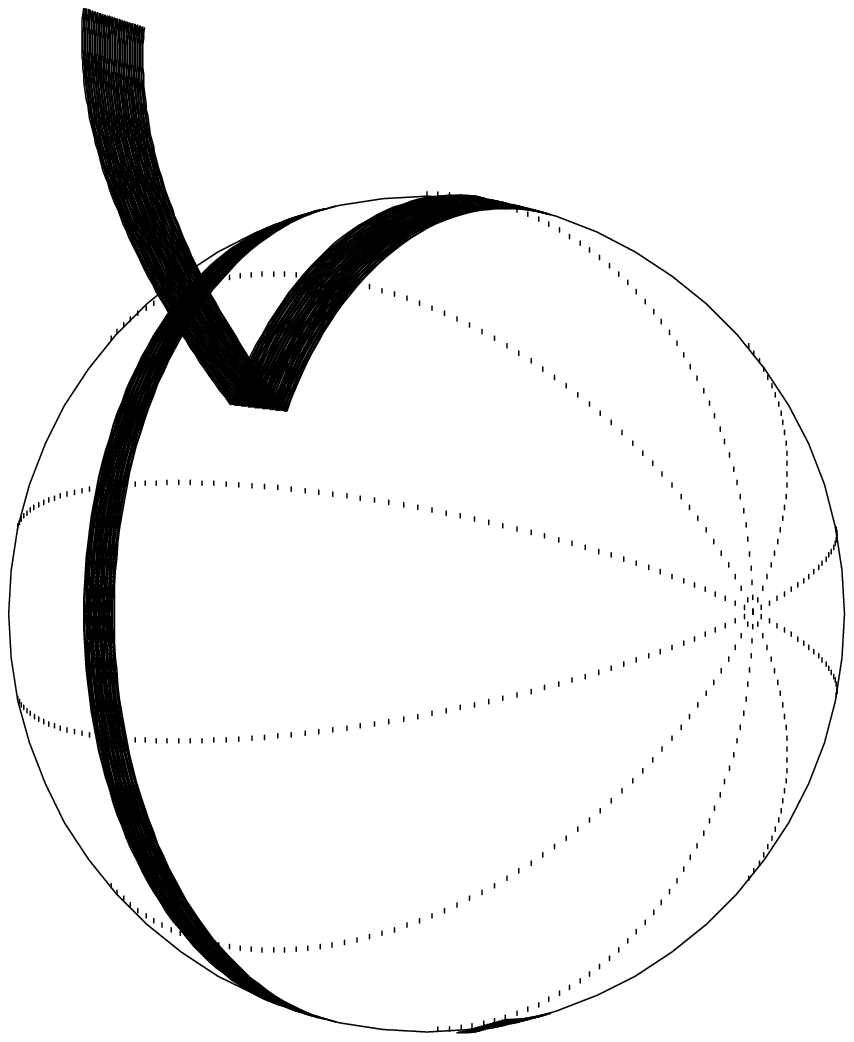}}
\centerline{Figure 5}
\centerline{\it Unwrapping the surface near Vancouver onto a flat surface.}
\bigskip

I must at this point insert a few warnings. As I stated earlier, an
analogy will often contain features which are not applicable to the  system
being modeled. This  holds true here as well. In the lenses I used to make
up
the space--time near the earth,  the time  decreases when
we get further
from the earth (i.e., the distance across the lens, and thus
the time as measured by a clock, is less far away than nearby).
However, the Vessot experiment showed that clocks far from the
earth run faster not slower. Had we constructed our lenses so as
to have them wider far from the earth than near the earth objects
would have been repelled rather than being attracted to the surface
 of the earth as represented on the beach ball. HOwever the beach
ball analogy also assumes that the notion of distance in time is
identical to the notion of distance in space. In particular it
assumes that distances in time and space obey the usual Pythagorian
 formula, that the square of the distance along the hypotenuse of a
triangle is the sum of the squares along the other tow sides.
 However in General and Special Relativity space time is not strictly
analogous to space. In particular,
the Pythagorian theorem for time contains a minus sign--- the distance
in
space--time looks like $\Delta x^2-\Delta t^2$ --- The distance
along the hypotenuse equals the difference in the squares of the
 other two sides, if one of the sides is in the time direction.
 These two effects, the difference in sign and the difference in rate
of flow between our model and in reality, cancel so that
 both  the analogy and  reality have the feature that
thrown objects return to the earth.

 Secondly, the
feature that time reoccurs on the beach ball  --- i.e., by going forward
in
time you get back to where you started--- is not shared by the structure
of time near the earth. It is however interesting that once we have allowed
time to become dynamic, once we have allowed time to flow unequably, situations
like space-times with closed time-like curves (time travel) are at least
a theoretical possibility. The past five years have seen an explosion of
interest in such situations in General Relativity\cite{TimeLoops}.

The third feature of the
model, that at a sufficient spatial distance from the earth the flow of
time goes to zero (the lens came to a point at the poles
of the beach ball) is inaccurate  for the earth and is actually also
unnecessary in the model. It would have been
 better to use lenses whose sides became parallel far from the
 earth, with only a
slight thickening near and through the earth.
 Such a model would also have allowed a demonstration of escape velocity.
On
the surface obtained by gluing a number of identical copies of this better
lens
shape together, straight lines which have a sufficient  slope at  the
 earth's surface do not return, as they always do on the beach
ball, but continue to
spatial infinity. This added realism would however have been at the expense
of
 a greater difficulty in seeing what the straight lines are in the model.

I have  discussed the role of time in General Relativity at such
length and in such an elementary fashion, in order to emphasize the
radical nature of change wrought by Einstein in this theory, and to emphasize
that the revolution in the nature of time is not
 simply some  abstruse and technical  change, as
is often claimed for General Relativity. Rather the change is
 simple and fundamental. General Relativity is not simply another
rather complicated field theory, but changes the very nature of physical
explanation (what is gravity?) in a way totally unexpected and still largely
unappreciated by the general populace. We shall see furthermore after the
next section that this change of the role of time in General Relativity
has in fact led to a reintroduction of precisely the opposite of Newton's
definition of time, namely the conception he ascribed to the common man
(and which is today a conception utterly foreign to the common man).

\section{ Quantum Mechanics and Time}

Time is also an important player in the theory of quantum mechanics, but
it
is in many ways a very different player here than in General Relativity.
It
does not itself become a participant in the action, but compensates for
this by
assuming a much more important role in the interpretation of the theory.

I will identify and discuss, at varying lengths, four places
 in quantum mechanics
where time plays a crucial role. Two of these are in the area of what
 is usually called
 the dynamics of the theory, and the other two are
in the area of the interpretation of the theory.

I) A theory of physics operates by elevating certain features
 of the physical world to the status of fundamental features.
These features are modeled by certain mathematical structures,
 and the theory delivers its predictions and explanations in
terms of those structures. One of the key features in physics
 is that these fundamental structures are taken to be the same at all times.

At each instant of time there is a set of fundamental dynamical variables,
the fundamental quantities in
terms of which the
physical nature of the world is described.
In quantum mechanics these are represented mathematically as
linear operators
on some Hilbert space. For example, if our world
 of discourse consists  of some
single particle, the fundamental variables are position and momentum.
These are represented by Hermitian operators, and all  physical
attributes are assumed to be represented by functions of these operators.
Furthermore, as long as one's world of discourse remains that single particle,
these same variables will maintain their fundamental role.
On the other hand, if our world of discourse is a quantum field
theory,  (e.g., electromagnetism), these physical attributes
will be the value of the field at each
spatial  point, together with the conjugate momentum for that
 particular field value.

It is one of the peculiar assumptions
 in physics that for any physical situation,
the fundamental
 set of physical attributes in terms of which we describe that situation
are
the same at all times. The universe of discourse does not change
 from time to time. This is of course
 very different from most other fields of human endeavour, in which any
attempt to pretend that the essential entities in terms of which the world
 is constructed are always the same would be silly. Institutions fail and
disappear, department heads retire.
  In physics, on the other hand, one believes that  the physical world
can
 be described at all times by the same physical structures,
 and that the changes
in the world are simply the changing relations between these fundamental
entities or changes in the values of their properties.

One can regard this as either an admission by physicists that
they will limit
their interest only to those situations in which such a
universality and invariance of
the fundamental attributes is accurate or as the claim that
all situations, no matter how complex
and how changeable, can be described by such a single unchanging
set of fundamental
attributes. The almost universal belief among physicists is that
the latter is the case, that at heart
there are some universal fundamental structures that can be used
 to describe any physical
process in the world.

Let me state this assumption in a slightly more technical vein.
To do so I will review the basic mathematical structure of
quantum mechanics. I would refer the reader who is not familiar
 with quantum mechanics to various books on the subject or to the
article by Shimony\cite{Shimony}. In
quantum mechanics, the basic entities used to model the physical world
are linear operators on a Hilbert space.  A Hilbert space is a collection
 of things, called vectors, which you can add together to get another vector
 or which you can multiply by a complex number to get another vector. I
will denote these vectors by $\ket{name}$ where `$name$' is simply some
symbol to name a particular vector. Thus linearity is expressed by the
statement that if $\ket a$ and $\ket b$ are both vectors, then
 $\alpha\ket a +\beta\ket b$ is also a vector for arbitrary complex
numbers $\alpha$ and $\beta$.
 These vectors also have associated with them a concept of
`inner product', designated by  $\braket ab$ which is a complex
number associated with any two vectors $\ket a $ and $\ket b$.
It is chosen such that $\braket aa$ is always real,
 $\braket a{b+c}=\braket ab + \braket ac$, and
 $\braket {\alpha a}{\alpha a}=|\alpha|^2 \braket aa$ for any
complex number $\alpha$. (Note that the vector $\ket {\alpha a}$
is used to designate the vector $\alpha\ket a$ and $\ket {b+c}$
designates the vector $\ket b +\ket c$.
The real number $\braket aa$ denotes the length squared of the
vector $\ket a$.)

These vectors in the Hilbert space do not  have any
direct physical meaning, but they do form an important element in
the interpretation of the theory and  in the establishment of the
 relation between the theory and particular realizations of
the theory in the real world.
The fundamental
 attributes of the physical world are the linear operators
 on this Hilbert space. An  operator is any function, which takes
 as input a vector in the Hilbert space,  and gives
 as output a possibly different vector in the same
 Hilbert space. If we designate an operator by the
 capital letter $A$, the we designate the result
 of the operation by $A(\ket b)$ or often by $A\ket b$.
These operators are linear if the result of the operation of the
 operator on the vector which is  the sum of two input vectors
 gives as its output a vector which is  the sum of the two
individual outputs, i.e., if $A(\ket b +\ket c)=A(\ket b) + A(\ket b)$.
In the theory there are two types of operators which are of
 special interest-- Unitary and Hermitian operators. Unitary
 operators are operators which do not change the length of
a vector. I.e.,, if the length of the vector $U\ket a$ is
the same as the length of $\ket a$ for all vectors $\ket a$,
 then the operator $U$ is unitary. Hermitian operators are
those for which their `expectation values' are real. The
 expectation value of an operator $H$ in a state $\ket a $
is the inner product between the state $\ket a$ and the
transformed vector $H(\ket a)$. In quantum theory all
 potential physical attributes of the world
  are modeled by
  Hermitian operators.

It would be at least  conceptually possible that  the Hilbert space, and
thus
the operators that act on the Hilbert space, could change from
time to time. One could imagine that certain things which were
 possible, which were measurable, at least in principle, at
one time,  did not even exist at some other time, that the set of all physical
quantities could be different at one time from the next. However this is
 not the case in quantum mechanics. The physical  variables, the
 set of operators and the Hilbert space on which those operators
 act,  are assumed to be the same at all times. As with classical
 physics, the change in the world that we want to describe or
explain lies not in a change to the
fundamental structures of the world, but in a change to the  relation
of these structures to each other or in the `values' that these operators
have.

This first role of time might appear trivial.
Time plays a role in designating  the set of fundamental
attributes in terms of which we describe
 the world.  Of course the physicist can decide to focus on some smaller
set
 of fundamental attributes of the world that is of interest at any given
time.
 But these changing simpler structures are not produced by
 some aspect of time, they are
produced by the changing focus of the physicist. The belief is strongly
held
that at heart there exists some one set of universal operators,
 some one global Hilbert space, which can be used (barring
technical difficulties) to describe everything in the world
throughout all time.

New theories may demand new assumptions.
The possibility exists that the world could change from time to
time in some fundamental way, not just in  detail.
 I will argue below that such genuine novelty may be needed in order to
describe the quantum evolution of the universe.

II) Having defined the mathematical structures used by the theory to describe
the world, one must then try to use them to explain the world. In
particular, one wishes to use the theory to explain the change we see
about us. Since the mathematical structure is
explained are constant, the explanation for change must be in
 terms of  changing
relations amongst the fundamental entities and changes in the
 the relation between the mathematical structures and the
physical world.   This explanation is done via the
equations of motion, equations relating the entities describing
 the world at one time to those describing the world at another time.

 I will work in what is termed the Heisenberg representation
 in which the identification of the operators on the Hilbert space with
 some physical attribute changes from time to time.
 There is an equivalent representation, the Schroedinger representation,
  in which the Hilbert space vectors transform over time
 but the identification
 of certain operators with physical entities remains constant. The two
are
equivalent in their ultimate predictions, but for various reasons,
 I favour the former. I feel that it makes  the
distinction between the formal, dynamical aspects of the theory and
the interpretative aspects clearer.

Which operator on the Hilbert space corresponds to which physical
 attribute of the world?  This identification can change from time to time.
 These changes form the essence of the dynamics of the theory,
 and are expressed in quantum mechanics by a set of equations of motion
which take the form of the
Heisenberg equations of evolution
$$ i \hbar{dA\over dt}= [A,H]\equiv AH-HA$$
 A is any dynamical operator representing some aspect of the physical
world, and H is a special operator in the theory,
the Hamiltonian, usually identified with the energy of the system.
It is by means of these
dynamical equations of
the theory, these changes in identification from time to time of
 the operators with the physical reality, that one hopes to encode
 the dual characteristics of the world as
 envisioned by physicists. That dual character is one of a fundamental
identity from one instant of time to the next (at all
times one can describe the world by the same set of operators),
 together with the possibility
 and reality of change
and transmutation, which is so much a part of the world around us.

III) Quantum mechanics arose out of, and encodes within its interpretation,
 a very
uncomfortable feature of the world,
that the world seems to operate
 on the basis of insufficient cause. Things just
 happen, without our being
able to ascribe any sufficient cause to explain
the details of what occurred. Given  two absolutely identical
situations (causes), the outcomes (effects) can differ.
  This feature  caused
Einstein so much intellectual
pain that he rejected the theory, even though he had
 been a key player in its foundation. It  is still one of the most
disconcerting aspects of the theory and the world.
 It seems to call into question the very
purpose of physics, and it lies at  the heart of the
disquiet  felt
 by even some of the best physicists\cite{Bell}.
 However, all   the evidence indicates that
 nature operates in accordance with quantum mechanics.
One's first reaction would be to say that
somehow at heart the universe surely operates with sufficient cause---
that
when we say that  identical situations produce differing results, it is
really that the situations were not
identical, but that there were overlooked features of the world which caused
 the differing outcomes.  However, the evidence is that this is not how
the
world operates, that God truly does ``play dice"\cite{Bell-Thm}.

How is this element of insufficient cause encoded in the quantum theory,
and how does time, the subject of
this paper, enter into this encoding?  As mentioned above,  each physical
attribute of the world is represented by an operator on the Hilbert space.
 But  physical attributes are
 not seen by us to be operators, rather they are seen to have definite
values
 and definite relations to each other. The position of my car is not some
operator which moves abstract vectors around but is some number,
 say twenty feet
in front of my house.  How
 can I relate the operator which represents the position of my car in the
theory, with  this number which represents the position of my car in the
world?

The answer is that  Hermitian operators have the  property that
there are certain vectors (called eigenvectors) in the Hilbert space for
 which the operation of the operator is simply multiplication by some
constant (called the eigenvalue).   Quantum mechanics states that the
 set of  numbers, the set of all  of the possible eigenvalues of the
 operator, is also the set of possible values that the physical attribute
 corresponding to the operator  can have. If my car can have any one
 of the real  numbers as a possible value for the
position of my car, the operator representing the position of my car
must have an eigenvalue corresponding to each of those real numbers.
The vectors of the Hilbert space are now used in the following way. If
 one of the vectors in the Hilbert space represents the actual state of
the world, then the theory tells us what the probability is that,
given the state of the world, the  value of the physical attribute
 corresponding to that operator takes some given value. I.e., the
theory does not tell us what the value of the attribute is, it
tells us what the probability is that it has some value.

 How do those values correspond to out experience of the physical
 world and our experience with those attributes that the system
has? In clasical physics any attribute of a physical system is
taken to have some unique value at all times. In quantum theory
the situation is more difficult. It seems to be impossible to
hold to the classical notion of each attribute having a value
 at all times. However there are certain situations, called
 measurement situations in which the attribute is taken to
 have some value, because it has been measred to have a value.
 Each operator representing a physical value
 has in general many eigenvalues, and thus the attribute has
many potential values.
The assumption is that at any given time, if a ``measurement"
is made of the attribute, one and
only one of these potential
values can be realised. The theory does not specify which
value will be realised, but rather gives probabilities for
 the various possibilities. Furthermore, these probabilities
 are such that the
probability of obtaining two (or more)
 distinct values is  zero, and the probability of obtaining any one of
the complete set of values is unity.
{}From the definition of probabilities, if $a$ and $a'$ are
two separate possible values for an operator
 $A$, then the impossibility of having two separate values
 gives $Prob(a~{\rm and}~ a')= 0$.
This then leads to
$Prob(a~{\rm or}~a')= Prob(a)+Prob(a')$. Furthermore,
if we ask for the probability that one of the eigenvalues
is realized, we have
$$Prob(a~{\rm or}~a'~{\rm or}~...~{\rm or}~ a'{}'{}'{}'{}') =
 Prob(a)+Prob(a')+...+Prob(a'{}'{}'{}'{}')=1
$$
where the set is the whole set of all possible eigenvalues of $A$.

Again this feature seems obvious, but time enters into this
statement in a crucial way. The statement that one and only
 one value is obtained is true only at some given specific
time. It is simply not true without the
temporal condition `at one time'. If we do not specify time,
 my car can have many
different positions (and it did today). It is only at a single
 time that
 we can state that the car had
one and only one position.

This seemingly trivial fact is encoded into quantum theory at the most
fundamental level. The probability of an eigenvalue is given by the
square of the dot (or more usually called the inner) product between
 the two vectors, or in symbolic terms,  the probability that the operator
 $ A$ has value $a$ is given by the square of the dot product
 between the eigenvector $\ket a$ associated with the eigenvalue $a$,
 and the state vector of the system, written $\ket\psi$, so that

$$Prob(a)=|<a|\psi>|^2$$. The statement that only one value can be
 realized leads to
$$Prob(a {\rm or} a')= |\braket a{\psi}|^2 +|\braket {a'}{\psi}^2$$
and that at least one value must be realized gives
$$Prob(a~{\rm or}~a'~{\rm or}~...~{\rm or}~ a'{}'{}'{}'{}')
=| \braket a\psi|+|\braket {a'}\psi|^2+...
+|\braket{a'{}'{}'{}'{}'}\psi|^2=1.
$$
It is because of this physical demand that we can use the eigenvectors
of Hermitian operators
as the models for physical attributes--- The eigenvalues and eigenvectors
 of Hermitian operators have exactly this required structure.

The third role that time plays in the theory, then, is that, given some
physical attribute, that attribute  can take one
of a whole range of values, but at any
 single instant in time  it can take at
 most one of those values (the values are statistically independent),
 and it must take at least one of those values (the values are complete).

As with the first property of time, this property
 seems almost to be trivial. It is at least very
difficult to see how it could be changed without
 completely altering the structure of quantum mechanics.
 Furthermore it would seem to
reflect a fundamental attribute of the real world.
 However, as I will argue that quantum gravity may require such a change.

IV) The fourth role that time plays is in setting the contingencies or
conditions for the predictions of the theory. Theories in physics are
designed to be broad. They are, especially if they are to be fundamental
 theories, designed to be applicable in any
and all conceivable situations. They are generic and
not specific, universal and not particular. But any
experiment, any experience of the world is specific
and particular. How can the theory be applied to these specific cases?

In classical physics, the answer lies in the `initial'
conditions. Although the theory itself is universal and
generic, the specific contingencies of any particular
situation can be encoded so as to make the predictions
of the theory specific and particular.
The theory itself identifies the dynamical variables in
terms of which one will describe any situation. The
equations of motion specify how the values of these variables
 at any one time are related to those at any other time. To
complete the picture, therefore, one must  specify the variables at some
one
 time. Given
 the values of all of the variables at one time, the values at any
 other time are completely determined by the theory through
the equations of motion.  This specification
of the values of all of the variables is given the name in classical physics
of
`initial conditions'.  Although the term `initial' is used,
there is no need that
 the values be specified at a time earlier than the time of
interest, or even that the variables all be specified at one
 time. For example, instead of specifying the values of the
 position and momentum of a particle at one  time, one
can specify the position at two separate times.
Both have the effect of completely particularizing the theory.  Having
 specified these initial conditions, the theory provides a complete
model for the world in some particular situation and for all times. Any
new information gleaned about the system, which takes the form of finding
new values for some variables at different times,
 must either agree in particular with the values predicted by
 the theory, or the theory must be thrown out as a model for
the physical situation. This is the situations which led to Laplace's famous
statement
\begin{quote}
... An intelligence knowing all the forces acting in nature at a given
instant, as well as the momentary positions of all things in the universe,
 would be able to comprehend in one single formula the motions of the
 largest bodies as well as the lightest atoms in the world, provided
 that its intellect were sufficiently powerful to subject all data to
 analysis; to it nothing would be uncertain, the future as well as
the past would be present to its eyes.\cite{Laplace}
\end{quote}

This is not true for quantum mechanics, although three
hundred years of classical physics still exerts its views
 on quantum theory, and quantum mechanics  is often treated as though it
 were cast in the same mold.

As mentioned, quantum mechanics is a theory of insufficient
 cause. The complete specification of the theory at one instant
 of time is not sufficient to completely specify the outcomes
 of any experiment at  all other times. Some things will `just
 happen'.
 As a result the effect of the setting of the conditions on the predictions
 of the theory are much more subtle, complex and profound than they
 are in classical physics.

Let us begin with the simplest text book case.  Let us say that at
some time $t_0$, we know{\cite{know}} that the dynamic
 variable $A$ has the value $a$.
 As I stated, the value $a$ must be one of the eigenvalues of $A$
and has associated a vector in the Hilbert space called
 the eigenvector, which I will write $|a>$.  Since we know the
 physical variable
 $A$ to have value $a$, we need that the probability that it has value
 $a$ be unity, and the probability that it have value $a'$
 different from $a$ be zero (it can by property III have
 only one value at a given time.)  As stated previously,
 the probability of having a value $a'$ for some vector in the
Hilbert space $\ket\psi$ is given by the square of the dot product
 between the eigenvector $\ket{a'}$ and the state vector $\ket\psi$.
It is one of the fortunate features of Hermitian operators, that
 $\braket{a'}a=0$ for eigenvectors associated with different eigenvalues
 of the same operator. If we choose the state vector to be $\ket\psi=\ket a$,
we will precisely encode the belief that we know the value of property
 $A$ to be $a$. The state vector for the system is thus the way we
 have of encoding the
conditions under which we want the theory to deliver answers to us.
In particular we choose the state vector to encode the
information that we know that some physical property has some definite
value.

Now, for any other operator $B$, at a later time say,
 we can calculate the probability that the physical variable
 represented by $B$ has value $b$ by the square of the dot product
 $|\braket ba|^2$. Note that this does not in general lead to the
statement that
 the system has some value $b$ at that later time, as it would
 in classical physics. It leads to the statement that there is
 some probability that it has the value $b$. However, what value
 it will actually be found to have if it were measured is unknown.
It could have any of the allowed values
(i.e., those  with non-zero probability).
 The actual value `just  happens', and the theory can give no further
 cause as to why that, and not one of the other possible values, obtained.

This procedure has much of the flavour of classical physics. Knowing
 that $A$ has value $a$ allows us to assign as an `initial conditions'
the `value' $|a>$ to the wave function of the system. Future predictions
 now use this `value' of the state vector. That the `value' is actually
a vector in the Hilbert space rather than  the
assignment of the value $a$ to some specific variable could be seen as
a
difference
 in detail rather than essence. The theory differs from classical physics,
 but it would seem in this description to fall into the same `initial
 condition--- equation of motion' framework as classical physics. This
is especially true if one works in the Schroedinger, rather than the
Heisenberg representation of quantum mechanics where the dynamical
 evolution is encoded in a time variation of the state vector or
the theory rather than in a changing identification of operators
with physical attribute. The Schroedinger equation looks like
$$i \hbar {d\ket\psi\over dt} = H\ket\psi,$$
and the specification of the condition now looks like a specification
 of the initial conditions for $|\psi>$. This fact together with the
three hundred year involvement with classical physics has helped to
confuse students into thinking that quantum physics really is
little different from classical physics.

 The differences between the two are not
 simply matters of detail, however, but are really a matter of
essence. This becomes clear if we now ask a further
 question. In addition to knowing that the value of $A$ at time $t_0$ is
 $a$, I now also know that at the time $t_1$ the value of $B$ is $b$. How
do I encode this additional information into the theory? In classical physics
you don't.  The additional information is either redundant
(in that it does not alter the predictions which could have
 been made using $a$ alone), or it is inconsistent, in which
case the theory must be scrapped. (I am  assuming that the
knowledge of $a$ was complete,
in that it specified the value of all dynamic variables in the
classical theory. If not, the additional knowledge of $b$
could further refine one's knowledge of the `initial conditions'.
 The knowledge of $b$ could not make something that was impossible
 knowing $a$ alone into something possible with the extra knowledge.)

In quantum mechanics the situation is very different. The
knowledge of $a$ predicted only a probability for the various values
that $B$ could have. The additional information that $B$ had value
$b$ is therefore certainly not redundant- the knowledge of the actual
 value is obviously a much stronger piece of information
 than merely a list of probabilities. It is also in general not
inconsistent-- anything with a non-zero probability could after all
 occur. Since the knowledge of $b$ and $a$ is stronger than the
 knowledge of $a$ alone, how is this  new stronger piece of information
incorporated into the theory for the prediction of the values of
 some other variable  $C$ say? The elementary answer is that if $B$ is
 later than $A$, then $B$  supersedes $A$. I.e., for all measurements
 made after that of $B$, one uses the wave function $\ket b$  as $\ket\psi$
rather than $\ket a$.

The process of replacing the knowledge of $A$ by the knowledge of $B$
is traditionally called the ``collapse of the wave packet". It has
caused much confusion in the literature. In the Schroedinger
representation, in which the wave function changes both
due to the dynamic evolution and due to such `collapses',
this change in the state due to a change in knowledge has called forth
much, in my opinion, misplaced speculation
about the dynamics of such a `collapse'.  The collapse is not dynamics.
It is the incorporation into a theory of insufficient cause of new
information, of new conditions,  not contained in the old information.

This rule, that later knowledge supersedes earlier, has also led to
 comments that quantum mechanics is, in some sense, inherently time
 asymmetric. After all, the latter supersedes the former, not the
other way around. However, the appearance of time asymmetry is
due to the fact that the
 question being asked is inherently time asymmetric. The latter
replaces the former {\bf if} one is asking questions about the
system at a even later time. The latter does not supersede the former
in other cases. To highlight this point let us ask
a different type of question. Given that I know that $A$ has value
$a$ at time $t_0$, and that $B$ had value $b$ at a later time $t_1$,
what are the probabilities that $C$ had value $c$ at an intermediate
time $t_2$  between $t_0$ and $t_1$?.

As specific examples are often more comprehensible than general arguments,
I will present a specific example, but the conclusions drawn will  have
 general applicability.  As always in the field of
 interpretation, one tries to work with as simple a
 system as possible so as not
to obscure the essential point with a forest of technical
 detail. I will therefore
take the ubiquitous spin $\half$ system,
 with the most trivial of Hamiltonians,
namely zero. The dynamics are therefore trivial in that
 nothing ever changes---
the operators at different times are related to each other
 by being the same at all
times. The dynamical equations of motion  are
$${dA\over dt}=0$$
A spin $\half$ quantum system has the peculiar property that the value
of the
spin in any direction can only take one of two values, namely $\pm \half$.
 We are going to assume that at some time, say one of 9AM, 10AM and 11AM,
we know that the physical system has a value for the x component of the
spin of
$+\half$, Similarly at  another one of those three  times, we know
that the system has a
value of $+\half$ for the y
component. The question we ask is ``What is the value of the
component of the spin
along an axis between the x and y axes making an angle of $\theta$
to the x axis at the third time?"
 In classical physics, the answer is
simple and straightforward. Because the dynamics is trivial, we then know
that
the x and y components of the spin are independent, and  both had value
 $\half$ at all times.  Furthermore since
spin is a vector, the midway component, let me call it
$S_{\theta}$, will just be an appropriate sum of the
 two known vector components, and must have a value of
${\sqrt{2}\over 2}\cos(\theta-45^o)$.  Having specified
the values of $S_x$ and $S_y$ the values of all intermediate
 components are specified uniquely at all times.

In quantum mechanics on the other hand,
the situation is more complicated. There are in principle six different
answers, depending on the times at which the system was assumed to  have
had
those values in relation to the time about which we are asking the question.
Let me write $S_aS_bS_c$ to designate that the condition of  $S_a$
having some value
is for 9AM, $S_b$ for 10AM and $S_c$ at 11AM. The various
 possibilities for the temporal
order of the conditions are the six
permutations
$$ a)S_xS_yS_{\theta}~~~~~~~~~~~b)S_y S_xS_{\theta}~~~~~~~~~
c) S_x S_\theta S_y
$$
$$ d) S_yS_\theta S_x~~~~~~~~e)S_\theta S_x S_y~~~~~~~~~f) S_\theta S_y S_x$$

I will concentrate on the first four of these. Although one can say something
 also about the last two cases, the potential controversy would hide the
point
I am trying to make. In the first two cases, the answers quantum mechanically
are different, while classically they must be the same. In the first
the value of $S_y$ supersedes
that of $S_x$ in determining the probabilities for the two possible
outcomes of $S_\theta$. In the second the value of $S_x$
 supersedes that of $S_y$. In each case
9AM condition is irrelevant, because the 10AM condition
completely supersedes it. One says that the state of the system is the
state
determined by the 10AM condition,
i.e., is the ${+\half}$ eigenstate of $S_y$ and $S_x$  respectively, and
the
 prediction for the probability that $S_\theta$ has value $\half$ at 11AM
is
${\half}(1+\sin(\theta))$ in case a, and $\half(1+\cos(\theta))$
 in case b. Note
that these are not the same as each other.

The cases c and d both
 also have  unambiguous answers in quantum mechanics, and both are identical.
The probability that $S_\theta$ will have value $\half$ is
$$P_\theta = {1+\cos(\theta)+\sin(\theta)+\cos(\theta)\sin(\theta)
 \over 2(1+\cos(\theta)\sin(\theta))}
$$

This probability function has  at least one peculiar property.   We see
 that the probability of measuring $S_\theta$ to have value $\half$ is
 unity (certainty) both when  $\theta$ has the value zero and when
 $\theta$ has the value $90^o$. I.e., the probability
 that one will measure $S_x\equiv S_0$ at the intermediate  time to
have value $\half$ is unity and the probability to measure
$S_y\equiv S_{90}$ to
have value $\half$  is also unity.  It is however easy to
prove that there exists no state vector whatsoever
in the Hilbert space of this
spin $\half$ particle which could give this result. I.e., there
exists no $\ket\psi$ such that
$|\braket {+\half,x}{\psi}|^2=|\braket {+\half,y}{\psi}|^2=1$.

 We learn  from this example that the temporal ordering of the
conditions that one places on the question that one asks of the
 theory are crucial
to obtaining answers from the theory. Unlike  classical mechanics, the
 conditions cannot in general be mapped back onto
 initial conditions. Time, and in particular temporal ordering,
is needed in a crucial way not just to set up the dynamical relations,
but also to extract
sensible predictions from the theory.

Another  lesson we can learn from this example is that quantum mechanics
does
not have a inherent time order to it. If it had, one would have expected
the answers to conditions c and d to differ. After all the
temporal order is completely
reversed in the latter with respect to the former. However the predictions
are
identical. The fact that a and c differ, even though just the temporal
order of
$S_y$ and $S_\theta$ have changed is no surprise since the condition
on $S_x$ at 9AM in both cases  ensures that these are not the time reverse
of
each other.

This is of possible relevance to the discussion that Roger Penrose
gives in his book,
``The Emperor's New Mind". He argues that quantum mechanics
itself implicitly contains a time ordering, that the
specification of the conditions leads to a clear and natural
 time ordering. The relevant section occurs on page 357,
where he describes an experiment in which a lamp is placed in front of
a half
silvered mirror. One now places a ``photon" detector
 just in front of the mirror, to
detect when the lamp sends a photon toward the mirror, and a
detector behind the
 mirror. One finds that although the probability is only $ \half $
that the second
one will have detected a photon when the first one does, the probability
is
unity that the first one will have when the second one is found
 to have detected
one. He then adduces this as an argument in favour of the
position that quantum
measurement contains within itself an arrow of time
(as well as a time ordering.)
I.e., the placing the condition on the photon exiting
and asking for the probability
that the photon entered is not the same as placing the condition on the
photon
entering and asking the probability that it exited.

\epsfysize=3in
\centerline{\epsfbox{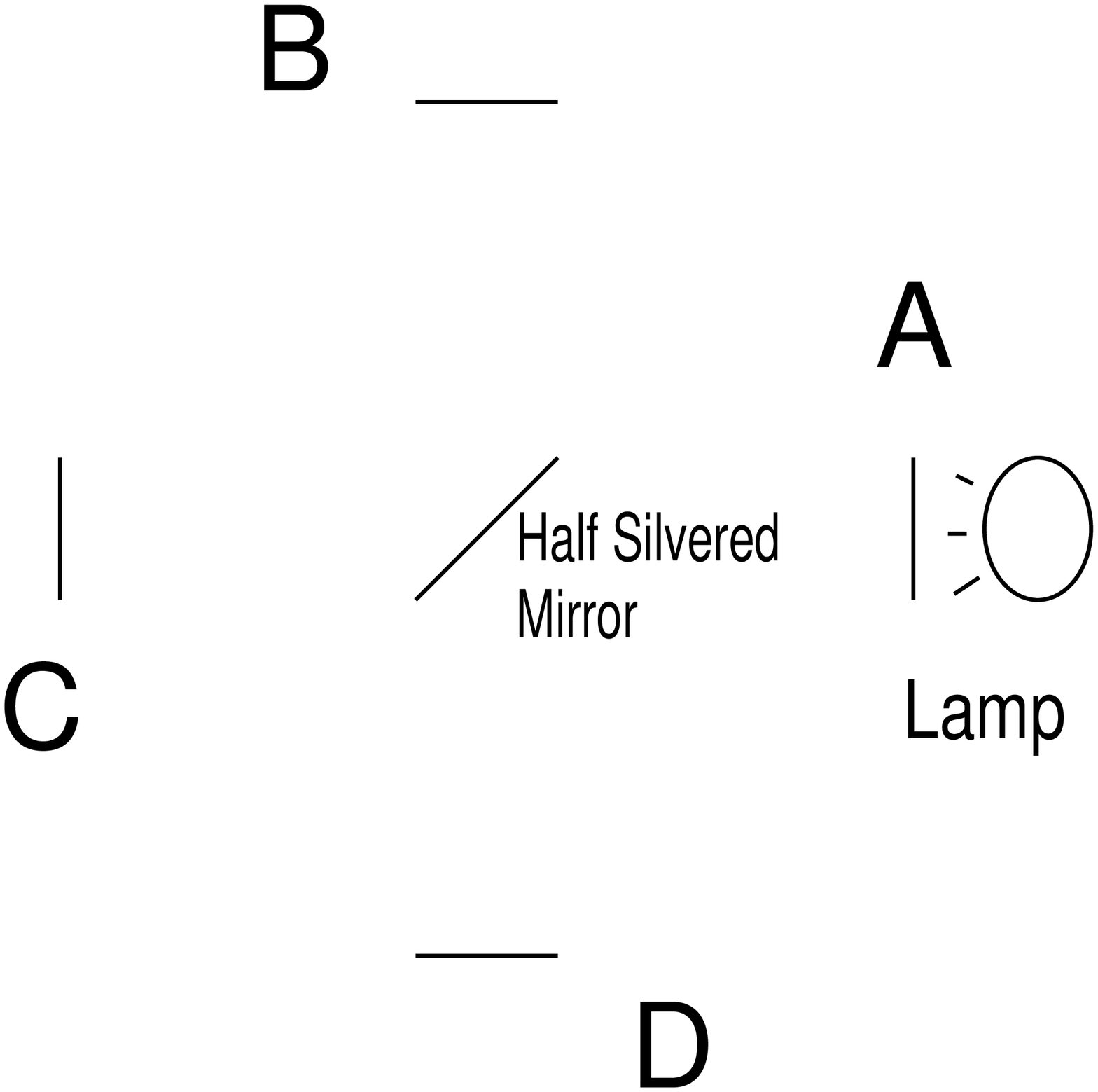}}
\centerline{Figure 6}

\centerline{\it Arrangement of the detectors and the half silvered mirror.}

Let me rephrase the problem posed by Penrose in a slightly different way.
Consider a half silvered mirror, with non-destructive photon detectors
placed at its
inputs A, B, C, D (see figure 6). Each detector detects the
presence of a
photon but lets it pass by unimpeded. Suppose we come across the notebook
 of an
experimentalist in which she has carefully
 recorded the readings at all  the ports.
 As expected, anytime either A or B gave a
reading, so did C or D. Furthermore, A and B
 never gave reading together, nor did
C and D. Beyond these facts however, the
readings turn out to be slightly  peculiar.
The notebook never
reported B as giving a reading. It seems that
 each photon  always passed through port A. The
 results for the C and D ports
were however roughly equally split, in that 50\%
went through C and 50\% through D. The
experimentalist furthermore stated on the page
that all four detectors were working
and that all readings in which at least one
detector fired were recorded. Do these results
 imply a time ordering  in and of themselves.?
Does this time ordering imply that the time
ordering is an inherent part of quantum mechanics?
  The naive answer is that the photons
 must have come in through port A and exited through
either of ports C  and D. We know that
there is no quantum state which would allow a photon
to come in through port C and exit, with unit probability,
 through port A,
while states exist for which the photon could enter port A
and exit with 50-50 probability through port C or D.  Does
this imply a time-asymmetry  in  quantum mechanics
   as Penrose implies?

The problem is that  there
 is something to be explained in either direction of time.
Ports A and B are completely symmetric. Why is there then an
 asymmetry in
the readings at the two ports, i.e., port A is the only one
 with any detections. The answer that Penrose would give
when we interpret A as the input port is
that there is a light bulb at
A and not one at B. The light bulb conditions the initial
state of
the electromagnetic field (i.e., we know that there are photons in the
initial state caused by the light bulb). However the measurement at A
coming later than the  conditioning by the light bulb supersedes the
effect of the light bulb, if we are interested in the outcomes at ports
C and D.
All we need to calculate the measurements at C and D is this knowledge
 that the photons came through A.
 If we now regard the system in the time reverse sense, where A is an
output port, there is again something to be explained, namely why did
 all of the photons exit through A. The photons come in at C or D with
equal
probability (which we would have expected {\it a priori}).  However none
exited at B. We cannot explain this from any initial conditions, because
the knowledge that the photon came in through C or D will supersede any
other initial knowledge. We can however explain it through a final
 condition, a condition which for some reason or another disallowed
 any recorded photons at  B.  Because we are now specifying
both initial (photon through either
 C or D) and final conditions, there will be no wave function which
encodes these conditions. However quantum mechanics can still make well
 defined predictions even in such an intermediate time case.

Now, one might   ask how one could arrange these final
conditions in any natural way.  After all we all know how
to set up the time reversed situation easily--- light sources
 are easily built to send photons into port A rather than
 port B. The reverse in which something disallows photons
from exiting from  B is not easily arranged.  However this
takes us into the whole realm of the usual statistical arrow
of time. Certain conditions are easy to arrange, and certain
 condition are very difficult to arrange in the world we live in. Cups
of
 tea cool not heat up when placed in a room-temperature environment.
 The difficulty in arranging the latter is not generally accepted
 as a proof that the fundamental laws of physics are time
asymmetric.  Similarly, the difficulty  of arranging an anti-light
 sink at B does not imply that there is anything fundamentally
 time asymmetric about quantum mechanics.

The fact that quantum mechanics itself does not pick out a
direction in time was recognized already over
 20 years ago
by Aharonov, Bergmann, and Lebowitz\cite{ABL}. They also
 suggested the formalism for handling
cases in which the conditions do not necessarily
precede the times at which
the predictions are to be made. That formalism has
 been independently rediscovered
a number of times since by others(including me),
 which illustrates the lack of impact that this fundamental
insight has had at least on the teaching of the subject.  It
 also illustrates the fact that the very different role played
by the conditions in quantum mechanics from that in classical
 mechanics has still not been instilled into the thoughts of most of
the practitioners of the subject.
In Appendix A I have outlined this formalism. For a more detailed exposition
I would refer the reader to the literature.

Aharonov and collaborators have recently been emphasizing the peculiarities
of
situations in which one sets conditions both before and after the
time at which one
wishes to discuss the possible outcomes of experiments. Because
 of the paradoxical
nature of some of these results, I will present
 one of their examples here\cite{Aharonov-spin}. This is
done to further reinforce the point I am making about the different
behaviour of
the conditions in quantum mechanics from the behaviour of initial
conditions in
classical mechanics and in addition illustrates the point that the
 subject of knowledge
and the relation of knowledge to physical measurements can be subtle in
quantum mechanics.

 The spin of a system in quantum mechanics is a vector $\vec S$, and represents
an internal type of angular momentum for the system. Angular momentum in
 quantum mechanics has a number of strange properties, one of which is
that
the angular momentum cannot take on any arbitrary value. The eigenvalues
for
 the operator corresponding to the total spin, namely $\vec S\cdot \vec S$,
take on only a range of values of the form $s(s+1)$, where $s$ is an integer
divided by two. I.e. the allowed values are discrete. Furthermore, if the
total
 spin is $s$ (which is the conventional shorthand for saying that
$\vec S\cdot \vec S$ has value $s(s+1)$), then any component of the spin can
only have values lying between $-s$ and $s$ and the value must differ from
$s$
by a whole number.
Consider some spin $s$ system At time $t=0$, we
measure the value of the spin in the $x$ direction (i.e.,
the x component of the spin), and find it to have
value $s$ (i.e., the maximum amount). At time $t_1$ we measure the spin
in the direction lying in the $x-z$ plane half way between the x and z
axes.
(i.e., we measure the operator $S_{45}={1\over \sqrt{2}}(S_x+S_z)$). This
measurement we carry out inexactly, to an accuracy of only of order
$\pm \sqrt{s}$. Finally
at time $t_2$ we measure the spin in the $z$ direction, and again find
it to be $s$. What is the probability distribution for the outcomes of
 the intermediate measurement of $S_{45}$? The answer turns out to depend
on exactly what one means by `measure inexactly', but there is a perfectly
well defined and acceptable meaning in which the answer is that the measurement
of $S_{45}$ gives a value of $\sqrt{2}s$, ie, about $40\%$ larger than
the maximum eigenvalue that $S_{45}$ has. Such an answer is obviously
silly if at the intermediate time we could imagine the spin system to have
some wave function which describes its state, since the mean value of the
probability distribution over the eigenvalues must be smaller than the
largest possible value. This strange, but true result arises from a conspiracy
between the fact that we are asking an intermediate time question (i.e.,
our conditions are not purely initial or final conditions, but rather are
mixed conditions), and the fact that the intermediate measurement is inexact.
Note that if the inexactness of the measurement is of order $\sqrt{s}$,
then for a sufficiently large $s$, $.4s>>\sqrt{s}$, i.e. the error is much
smaller than the deviation from the maximum value.

Let me put a bit more flesh onto the above bones. The initial and final
measurements are assumed to be perfect exact measurements. For the intermediate
measurement we will institute the requirement that the measurement be inexact
by coupling the spin system to another system which will be our measuring
apparatus. The measuring apparatus will be assumed to be a free particle
of infinite mass and zero potential energy- i.e. I will assume that
the free Hamiltonian of the particle is zero. (This is supposed to represent
say the dynamics of a massive apparatus pointer.) The coupling will be
taken
to be such that the interaction between the apparatus and the spin
system is instantaneous (i.e., a delta function in time) and is proportional
to the momentum of the particle times the spin. I.e. the full Hamiltonian
for the system is
$$H= S_{45} p \delta(t-t_1)$$
In the Heisenberg representation, we find that the dynamics of the apparatus
is given by
$$p(t)=p(0)$$
$$q(t)= q(0) + S_{45}$$
$$S_{45}(t)=S_{45}(0)$$
$$S_{-45}(t)= \cos(p) S_{-45}(0) +\sin(p) S_y(0)$$
The inaccuracy of the measuring apparatus will be introduced by assuming
that it is the value of $q$ which will be used to infer the measured value
of $S_{45}$. I.e., we will measure q exactly after the coupling between
the
particle and the spin has completed and use that value to infer the value
of
the $45^o$ component. To obtain the value of this component we must subtract
the
final value of $q$ from the initial value of $q$ since the coupling to
the spin
causes the value of $q$ to change. To mimic the inaccurate measurement,
I will assume that the initial value of $q$ is not exactly known, that
the
initial state of the apparatus is such that the initial $q$ has a spread
in
possible values over a range $\Delta q= \sigma \approx \sqrt{s}$.
I will take the initial wave function for q to be a Gaussian, centered
at 0 with standard deviation of $\sigma$.
$$\Psi(q)= {1\over \sqrt{ \pi\sigma}}e^{\left(q\over\sigma\right)^2} $$.
This measuring apparatus does behave like a proper apparatus should.
{\bf If} we assume that the state of the spin system is in fact an eigenstate
of $S_{45}$, the final probability distribution of $q$ values is just a
Gaussian, centered around that eigenvalue of $S_{45}$, with width $\sigma$.
I.e., the best estimate for the value of $S_{45}$ will just be the value
measured for $q$ with an uncertainty in the inferred value of $\pm \sigma$.
I have carried out the calculation for the situation presented in our problem
above, namely that $S_x$ had value $s$ before the measurement and $S_z$
had value $s$ afterwards for a value of $s=20$, and for various values
of $\sigma$. These results are presented in figures 7-10. In figure 7,
I have taken $\sigma=.5$. The intermediate apparatus is sufficiently accurate
to distinguish
 between the various possible values of the spin component (which we recall
must be separated by unity from each other.) The result is as expected,
a series of
Gaussian peaks centered on the expected possible values for the total
spin, with various heights representing the differing probabilities that
the spin has those different values.
\epsfysize=3in
\centerline{\epsfbox[50 170 624 744]{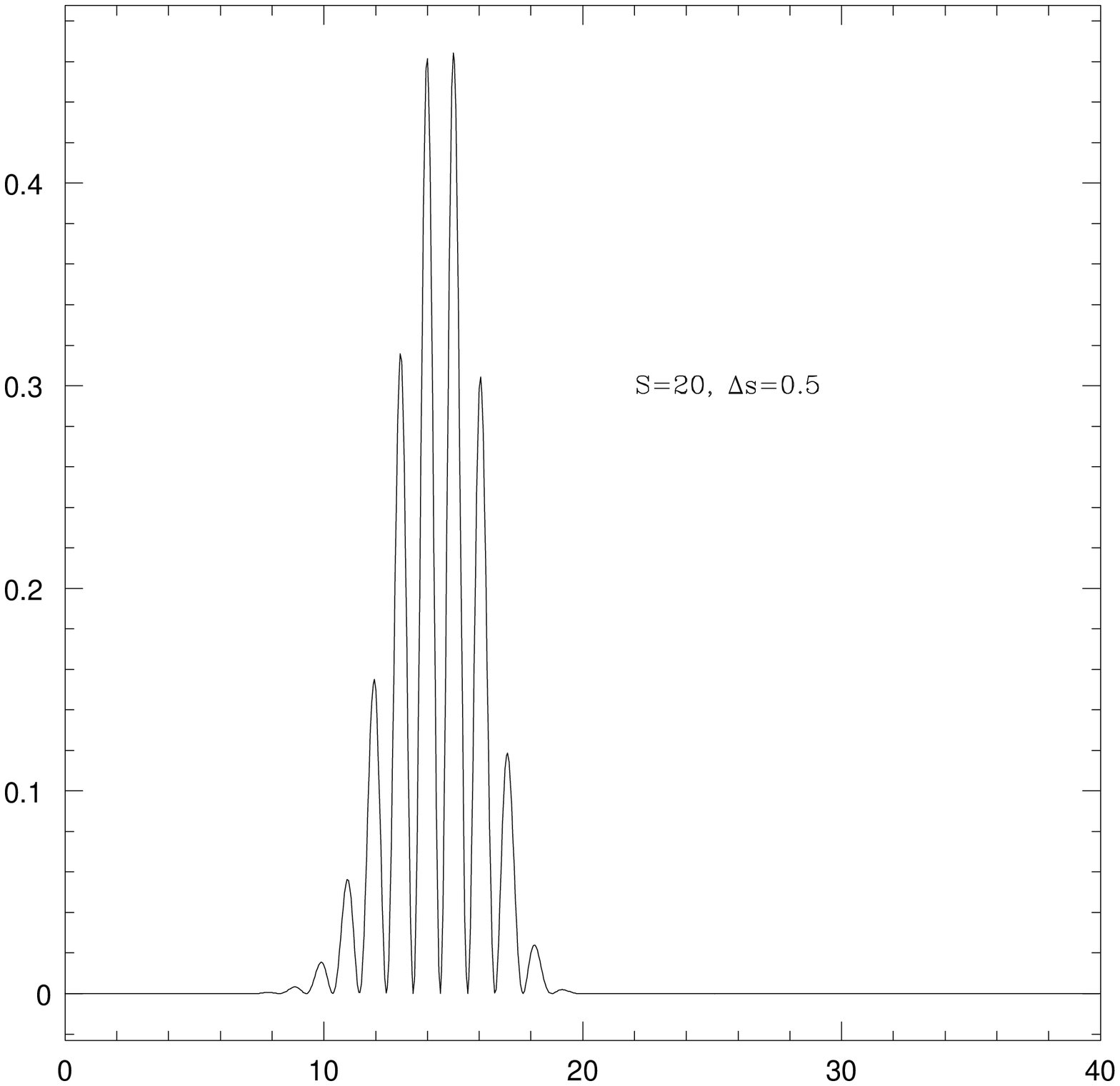}}
\centerline{Figure 7}
\centerline{ \it The Probability distribution for the pointer measuring
the spin with uncertainty .5--- Maximum spin=20}
\vskip .1in

In figure 8, I have increased $\sigma$ to 1,
and the measurement is not as accurate. There are still a series of peaks,
but these are no longer centered on the values we would have expected for
the
spin, i.e. they are no longer centered on the integers. In figure 9, with
$\sigma=2$ this trend away from the naive expectation has continued.
 The spacing between the peaks is definitely
greater than unity, and the peaks in the probability distribution have
begun to extend beyond
 the maximum possible value for the spin, namely 20. Finally, in Figure
10,
$\sigma$ has the value of 5. This is far too coarse to be able to distinguish
individual spins, which have a spacing of unity. However we notice that
the
expected value of $q$ is now about 28, 40\% higher than the
 maximum possible value
that the spin is supposed to be able to have. The measurement
 of $q$ would give
an inferred value of the spin higher than it could possibly
be. One might at this
complain that the measuring apparatus is poor, that it does not
 measure the spin
properly. For almost all normal situations it is ,however, a good measuring
apparatus for the spin. In all normal situations, in which one specified
only an
initial condition and not a final condition as well, the outcome would
have
been exactly what one would have expected- namely a sum of Gaussian peaks
centered around the allowed values of the spin. It is
because one has instituted
both initial and final conditions that the measurements have produced the
strange result of a value much higher than the maximum allowed value.
\epsfysize=3in
\centerline{\epsfbox[50 170 624 744]{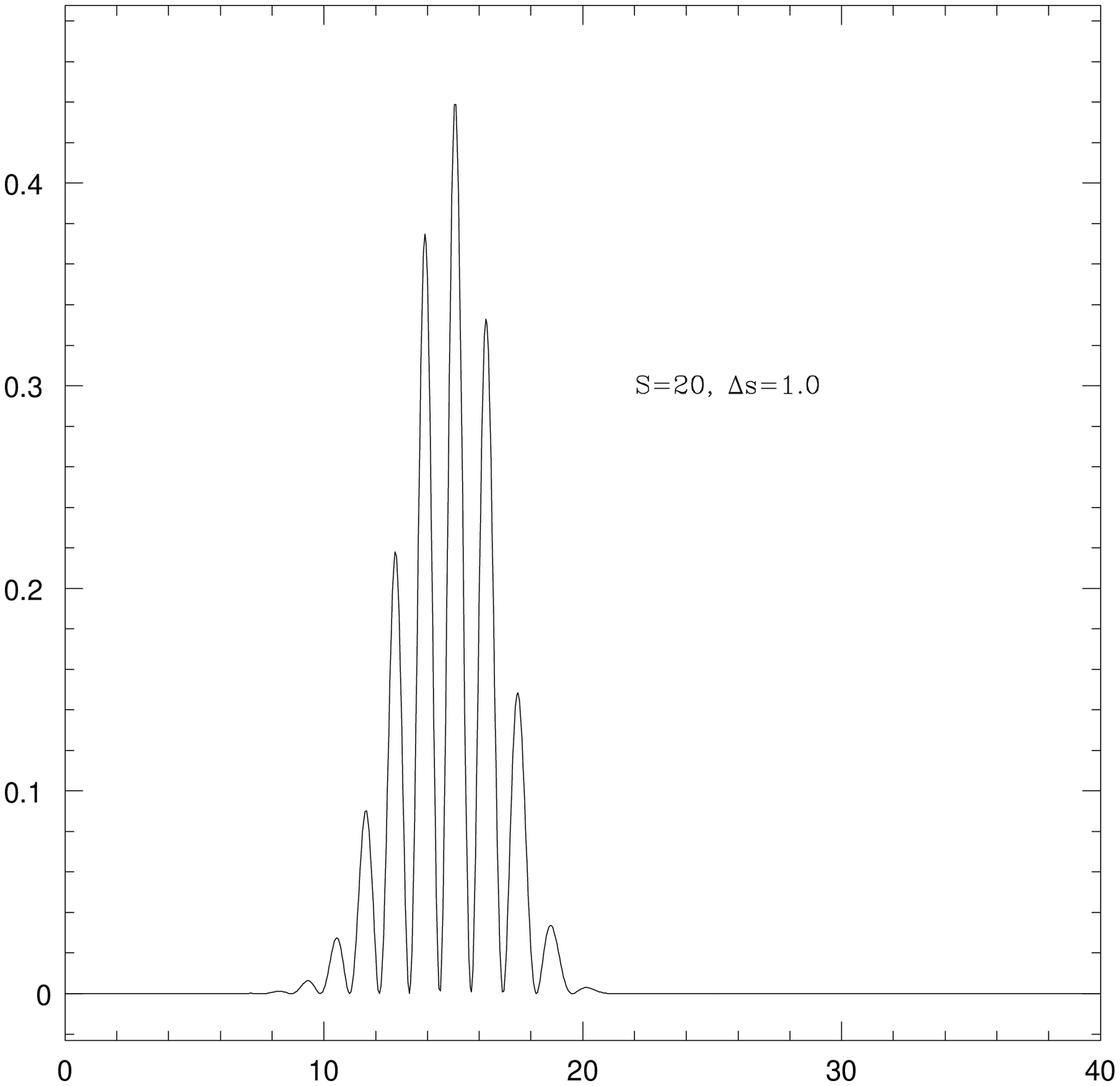}}
\centerline{Figure 8}
\centerline{ \it Pointer measuring the spin with uncertainty 1}
\vskip .1in
\epsfysize=3in
\centerline{\epsfbox[50 170 624 744]{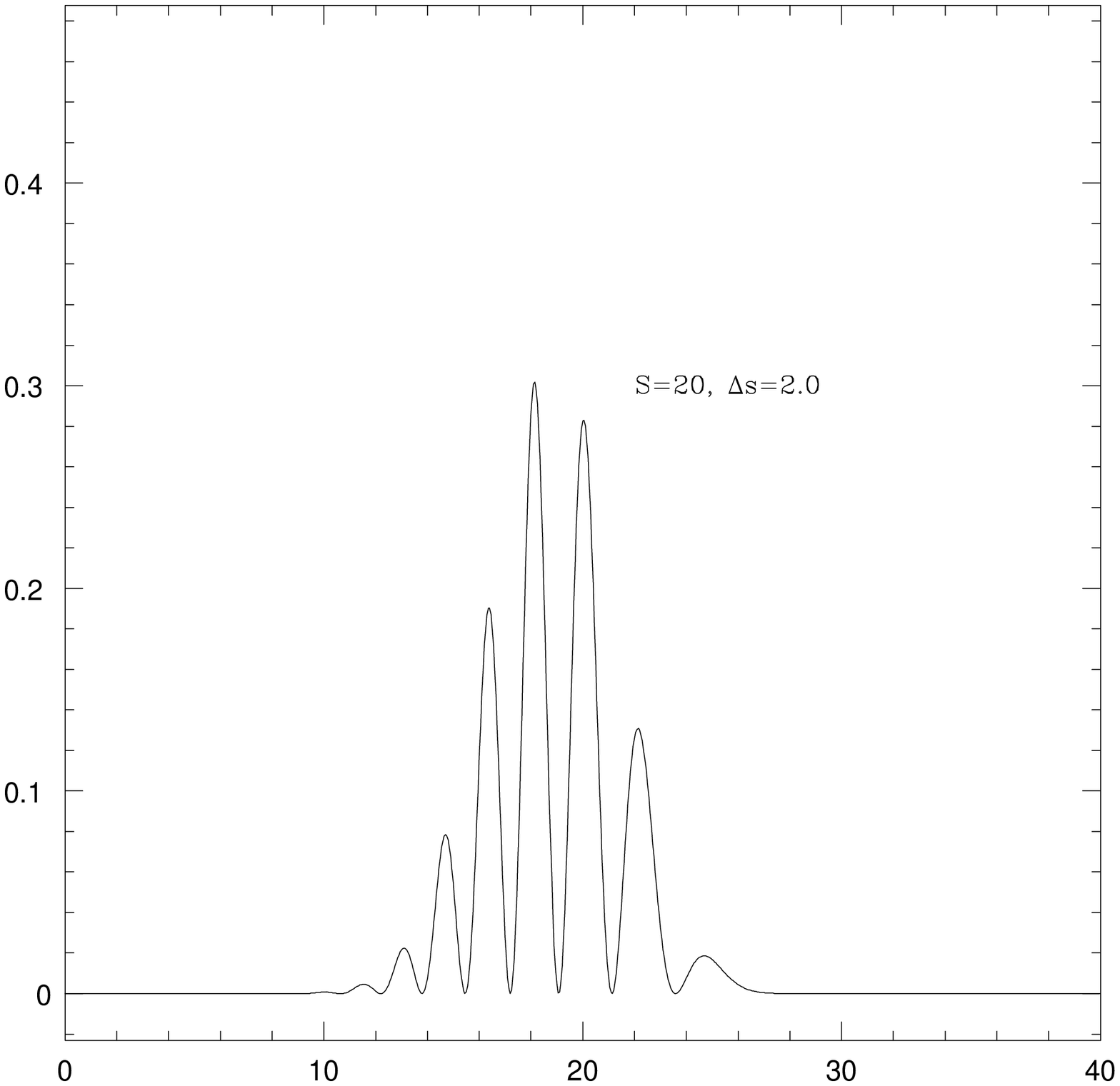}}
\centerline{Figure 9}
\centerline{ \it Pointer measuring the spin with uncertainty 2}
\vskip .1in
\epsfysize=3in
\centerline{\epsfbox[50 170 624 744]{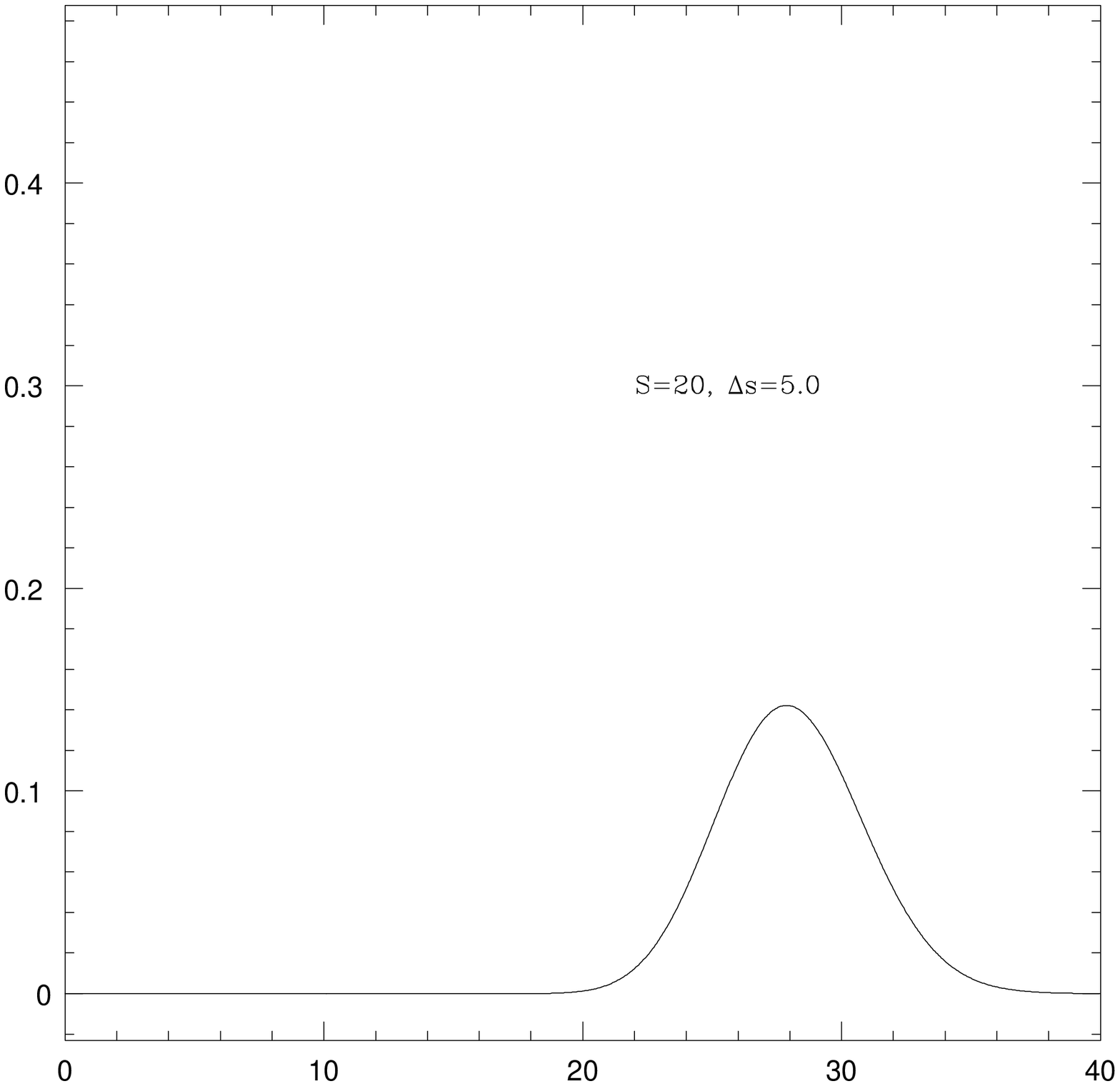}}
\centerline{Figure 10}
\centerline{ \it Pointer measuring the spin with uncertainty 5}
\vskip .1in

This is a wonderful example of the unusual nature of the
effect of the imposition of conditions
in quantum mechanics. The conditions do not reduce simply
 to initial conditions.
There are no initial conditions whatsoever which could
give the results of this
{\it gedanken} experiment. It is precisely because of the
 insufficient-cause nature of
quantum theory, that the temporal order of the conditions
is crucial. New knowledge
changes the results of the theory in a way completely
unexpected in classical
physics.

\section{Quantum Gravity}
In this section I will argue that attempts to quantize a theory
like General Relativity, in which time plays the central role, will potentially
meet with problems in all four of roles that time plays in quantum mechanics.
I say potentially, because we do not at present have a quantum
theory of gravity.
There may be some subtle way in which the difficulties can be
 avoided which we do not
at present recognize. The ultimate theory of quantum gravity
 may be so different
from our present notions that the role of time will not even be an issue.
(As an example, quantum mechanics itself was so
different from classical theory that
issues which arose in the latter were not even a question
in quantum theory.
Furthermore, the formalism of quantum mechanics itself
 suggested interpretations
which would never have occurred in classical physics.)
Despite the difficulty of discussing a non-existent theory,
it may be worthwhile
to look at the difficulties in the hope that a clear view of
of them may
suggest a solution to the problems as well. (See also the papers
\cite{Kuchar}\cite{UnruhWald}.)

I)
What is the Hilbert space and what are the operators of the theory?
And does one
expect that Hilbert space and the physical operators to be time
independent? I will
tackle the second question, and leave the first to the next section.
 One of
 the striking predictions of Einstein's thoery of gravity  was that
 not only was time intimately
involved in gravity but so was space. In particular
 one of the first set of solutions
to the theory were solutions in which space became
 dynamic, and the universe as a whole
grew in size. These solutions, first discovered by
Friedman, and hated at first
by Einstein, led to what we now know as the Big Bang
theory for the origin of
 the universe. In the popular imagination, the Big Bang
is like an anarchists bomb,
in which at some moment the universe, the cosmic egg,
 exploded,
and what we now see are the fragments of that explosion
hurtling away from us.
As usual, the popular image misses the most radical part
of the theory. It is not  that
there was some explosion in a preexistent space, but rather
 that the universe was born
very small, and as time went on, the universe created more
space for itself to live in.
The reason that we see the distant galaxies recede from
us is not that they are moving
away as the remnants of some inconceivable initial force,
 but rather that the distance
between us and them is increasing due to the creation of
new space
 between us and them at a more or less constant rate.
Because of this increase in the amount of space, anything
 in the universe is continually
being diluted. In particular, anything which now has some macroscopic
scale, very early on had a much much smaller scale
(a factor of about $10^{25} $ on the
 most naive assumptions, and a much larger factor
 difference if inflation actually
occurred).

Does this also mean that because there was less space early on there were
also fewer physical attributes that the universe had?
My suggestion is yes. Modern
physics sees the fundamental structures which make up the
universe as being fields.
Fields, like the electromagnetic field,  are entities that
have values at each point
in space. The number of different physical attributes of
 the world are thus
 related to the number of different field values at the
different points there
 could be. At first this would seem to suggest an infinite
 number, since there are
an uncountably infinite number of points in a classical space.
 However,
gravity itself would be expected to put a limit on the number
 of different values
that any field could have. If the field fluctuates at a
sufficiently small scale,
those fluctuations would have a sufficient energy to
collapse into a black hole. I.e.,
we would expect gravity itself to put a limit on the scale
of the fluctuations,
a limit which very naive estimates would put at scales of
 the order of the Plank
scale--- $10^{-33}$ cm or $10^{-44}$ sec. For example, one
 would expect that
field variations on scales smaller that these scales simply would not exist.
If the above is true (and it is very naive, although it gains a slight
amount
of support by realizing that the entropy of a black
 hole is directly related to the
number of distinct regions of Plank size that there
 are on the horizon of a black hole), then the number
 of distinct field values might
be finite in any finite volume of space, and that
 furthermore, the number would
decrease as the universe decreased in size. This would suggest that in
the
critical assumption behind condition I) would be wrong, that one should
rather
 describe the universe with a Hilbert space and a set of operators which
could
change in time, rather than be a constant. One of the real difficulties
is that
no-one knows how to implement a theory where the set of operators and the
Hilbert space change in time. One would also have to face the question
of what the
laws would be which would determine such changes. This idea has therefore
received
almost no study but remains a disquieting possibility.

II) Dynamics: In the model I presented for the unequable flow of time near
the earth,
I used the notion that the structure of the space-time near the earth is
essentially
constant throughout time. This allowed me to build up the space-time out
of
a number of identical pieces. It was also this notion of a time with respect
to
which nothing changed that allowed me to talk about an unequable flow of
time.
It was the distance notion of time, the time measured by clocks or other
 dynamical processes, which `flowed unequably' with respect to the time,
the `symmetry time',
with respect to which the space-time was the same from time to time.
In the generic situation one does not have any symmetry time, any time
with respect
to which nothing changes. Rather one must introduce an arbitrary notion
of time, what
is called coordinate time, with respect to which one measures change. Although
at
first this arbitrariness in the definition of time would seem to make any
definite
statements impossible, the techniques of differential geometry, which were
 developed at the turn of the century just before Einstein developed his
theory
by using those techniques, allow one to identify physical features which
are
independent of the arbitrariness of the choice of this coordinate time.
However, this arbitrariness has a consequence for the quantum theory. One
of
the consequences of the arbitrariness in the choice of coordinate time
is that
the natural variables in terms of which one would describe the space-time
are themselves partially redundant. There are too may natural variables,
and the theory
demands that some of them obey an equation known as a constraint equation.
Such
constraints are relations between the variables at one instant of time
rather that
relations between the variables at different instant of time, as in the
equations
of motion. These constraints are very difficult to implement quantum
mechanically.
The most natural technique, introduced by Dirac to handle any theory with
redundant
variables, leads to a restriction of the Hilbert space on which the physical
theory is to
be defined. This leads as well to a restriction on the operators. Because
of the
arbitrariness of time in General Relativity, these constraints on the natural
variables lead
to a Hilbert space such that the only natural operators defined on this
space are the
constants of the motion. Only those features of the space-time geometry
which
do not change from arbitrary time to arbitrary time seem implementable
as
operators on the Hilbert space. Thus the dynamical content of the theory
seem
to be trivial- nothing changes.

How then do we describe or explain the change that we experience in
the world around us? This is one of the questions that most bedevils any
putative theory of quantum gravity. Although many have thought about the
problem,
and some have felt that they have solved it to their own satisfaction,
there is no
generally accepted answer. Moreover I feel that all of the suggested answers
(including my own) have severe difficulties. How can change be described
in a
theory where the only valid physical quantities that do not change?

III) Probabilities:
One of the suggested resolutions of the problem mentioned above is to select
one of the variables of the unconstrained theory is selected as the time.
 The physically intuitive reasoning is that time in reality is an
 unobservable feature of the world anyway. What actually passes for
time is the reading on various and sundry pieces of physical apparatus
 called clocks. If you as a child are late for school, it is not because
your arrival at the school is late in relation to any abstract notion of
time. It is
rather that the reading on the face of your teacher's watch is later than
the reading
at which school was supposed to start. Note that this approach is in direct
contradiction to Newton's approach as stated in the quote which began this
paper.
Time, according to the proponents of this view is exactly the common view,
and
Newton's non-relationist  view is wrong.
The key problem with this approach is that it removes the foundation for
the third
aspect of time in quantum mechanics. At any one time, any variable has
one and only
one value. It is this which physically justifies the whole Hilbert space
structure
of quantum mechanics. But any real physical watches are imperfect. It can
be proven that
any realist watch not only has a finite probability to stop, it has a finite
probability
to run backwards. Now as long as the watch is simply the measure of some
outside
phenomenon, one could take these probabilities into account. If, however,
 time is {\bf defined} to be the reading on the face of the clock, the
question
as to whether or not the clock can stop or run backward is moot--- it cannot
by definition.

However the other question now raises itself, namely what basis  have we
for the
assertion that at a time a physical quantity can have one and only one
value.
At the same readings on a broken watch, a physical quantity can have an
arbitrarily
large number of readings. There is furthermore no reason why the probabilities
should add up to unity if the events are not mutually exclusive. They could
add up
to far more than unity, and still be in accord with probability theory
if they are
independent, mutually exclusive events. What is the mathematical structure
which
should be used if time no longer plays this third role of defining he sets
of
mutually exclusive and exhaustive possibilities?

IV)
We argued in part IV) of the last section that the setting of conditions
in quantum mechanics was
more subtle and rich than in classical theory. In particular, the setting
 of conditions is not equivalent to the setting of initial conditions.
Furthermore,
unless the experiments were structured so that the conditions always preceded
the
questions, the usual use of a state vector to encode the conditions was
inapplicable.
How does this apply to the attempts to quantize gravity?

One of the approaches is to regard the constraints as a sort of Schroedinger's
equation, called the Wheeler-DeWitt equation.  The solutions to this equation
 are then supposed to correspond to `wave functions of the universe'. If
 we furthermore regard, a la the previous section, one of the unconstrained
variables as the time, the equation is to represent the probability of
measuring various quantities at the `time' denoted by the value of the
variable chosen to play the role of the watch.

In this approach, one of the questions has been: how does one choose which
of the
possible solutions to the Wheeler DeWitt equation is supposed to represent
the
real world? One of Hawking's key contributions to the field was to suggest
that there was a natural choice for the `wave function for the universe'.
 He suggested that the `initial value' problem for the wave function could
 be solved in a natural way. The technique involves a trick. One can rewrite
the Wheeler DeWitt equation in terms of what is known as a path integral
in which one integrates the exponential of a function of the geometrical
 configuration called the action. If instead of using the action in which
the time component has the minus sign in the Pythagorean formula, one instead
uses the geometries in which the time component is treated in all respects
like a
space component, one can find a natural solution to the Wheeler DeWitt
equation.
This solution is obtained by performing the path integral over all geometries
which
are completely closed except for a boundary representing the geometry one
is
interested in. This approach is described in Hawking's best seller,
 {\bf A Brief History of Time}.

However in light of the discussion I gave in the section on quantum mechanics,
the
problem of setting the conditions in quantum theory is not a problem of
initial
conditions. In particular in the case in which the definition of time,
 of before and after, is problematic, one does not expect a wave function
to
properly encode the setting of conditions. Since the physical answers differ
substantially when the time ordering is changed, the lack of such a time
ordering
in the quantum theory of gravity makes this a critical issue. In what order
are the conditions to be set? If a wave function is not the right formalism
for
describing the theory, what is the correct formalism?

As should be obvious, the conflicting roles of time in gravity and in quantum
theory
have raised a number of difficult issues. Although the field seems no nearer
resolution
than it did forty years ago, it does seem that the effort to understand
the problems
has given us a much better understanding of the roles played by time in
both general
 relativity and in quantum theory.

One of the only people to try to come to terms with the above
difficulties in the formulation
of quantum theory for gravity has been Jim Hartle\cite{Hartle}.
In a series of recent papers he has been applying the consistent
 histories formalism (briefly sketched in the Appendix A) to a the
problems of Quantum Gravity. His attempts are still in an
 embryonic stage, but the formulation is one in which the
usual operators on a Hilbert space approach of conventional
 quantum mechanics is abandoned. Instead, starting from
 Feynman's `sum over histories' approach, he casts quantum
gravity directly in terms of histories of observation, histories which
need
not be tied to particular instants of time or locations in space.
It incorporates
the possibility for the
setting of conditions at arbitrary instants and ordering in time. It might
also
allow one to speak of situations in which the possible observables can
change
 in time, as in the concerns above regarding part I .  The exact nature
of time,
 of dynamics and change, has still to be elucidated in my opinion, but
it is possible
 that our present notions of time can arise from the theory as an
approximation.
 But to go into the details of his attempts would make this paper much
larger than
 it already is, and I will refer the reader to his papers instead.

{\section{ Appendix A}}
In this appendix, I want to introduce the formalism, developed by
 Aharonov, Lebowitz and Bergmann \cite{ABL}, by
Griffiths\cite{Griffiths}, by me\cite{Unruh}, and by
Gell-Mann and Hartle\cite{HGM} to describe quantum mechanics under conditions
in
which the conditions are set at arbitrary times and not solely before the
time of
interest. This formalism is used by all of the above people in
different ways, and Griffiths and Gell-Mann and Hartle have tried to use
it to
define a new interpretational scheme for quantum mechanics. I will not
go into the
details of that scheme here but  refer the reader to the literature.
 To develop the formalism,  we must introduce some notation. We will
again operate in the Heisenberg representation.
For any operator {\bf A} with discrete eigenvalue $a$, we can define a
Hermitian projection operator ${\bf P}_a$, such that
 $P_a^2=P_a$, such that it commutes
with {\bf A}, and such that it obeys $AP_a=aP_a$. To
use the language of Hartle and Gell-Mann,
we now define a "history" as a sequence of eigenvalues
 of various operators
at different times, ${a_1,b_2,...,z_n}$ where $a_1$ is
 an eigenvalue of $A$ at time $t_1$, etc., and $t_1>t_2>...>t_n$.
 They  define an operator
$$C_{a_1b_2....z_n}= P_{a_1}P_{b_2}...P_{z_n}$$.
This operator represents the successive projection onto the eigenstates
of the sequence of operators $A_1,...,Z_n$. Let the "vector" of eigenvalues
$${\bf v}=(a_1~b_2~...~z_n)$$
represent a history, so we can write
$$C_{a_1b_2....z_n}=C_{\bf v}$$
 Now define the ``decoherence" function
$$D({\bf v}; {\bf \tilde {\bf v}})= Tr(C_{\bf v} \rho_I \tilde C_{\tilde{\bf
v}})$$
where $\tilde {\bf v}$ is assumed to be another `history' of possible
outcomes of the {\bf same} sequence of operators as in {\bf v}. $\rho_I$
is the initial density matrix, which I will discuss below. $Tr$ denote
 the trace of the the operator (the sum of the diagonal elements of the
matrix).
We will divide the components of {\bf v} into two categories. The one category
is those values which we know--- i.e. those which we wish to use to set
the
conditions on the questions we ask of the theory, and the subset which
we do not know, i.e., those for which we wish to determine the probability
of that particular outcome. Let me designate the first subset (``known")
by $K({\bf v})$, and the second by $Q({\bf v})$. Then  one finds that
$$
Prob(Q({\bf v}))= {D({\bf v};{\bf v}) \over \sum_{Q({\bf v})}
D(\tilde{\bf v};\tilde{\bf v})}
$$
where
 the sum is over {\bf all} possible  {\bf v} obeying this
condition. Notice that we have chosen the same history in both slots of
the
decoherence function $D$ in order to define the probabilities.
It is worth pointing out features of this formula. If all of the conditions
imposed
occur at a time before any a question, then the above formula simplifies.
(I.e.,
 if $K({\bf v})$ all occur before $Q({\bf v})$, the above may be written
as
$$Prob(Q({\bf v}))= Tr(C(Q)C(K)\rho_IC\dagger(K)C(Q))/\sum_Q ...$$
We can thus replace the $\rho_I$ by a new $\rho= C(K)\rho_IC\dagger(K)$.
Now $\rho_I$ in the above is supposed to represent the initial ``density
matrix"
for the system. We note from this paragraph, that in reality the initial
 density matrix represents either some theoretical prejudice about the
truly
initial state of the system, or represents the cumulative effect of all
of those
conditions placed on the system at times earlier than the time in question.
We also note that we are free either to change the initial density matrix
in this way once
we set more a priori conditions, or to simply include the extra conditions
in the
formula as parts of the $K$ terms in the history. The first option, changing
the density
 matrix, is the process known as
`reduction of the wave packet' in the conventional
approach to the interpretation
of quantum mechanics.

It is useful at this point to mention that the Decoherence functional has
been made the center of a new interpretation of quantum mechanics. In my
presentation above, the results are all just a minor modification of the
standard interpretation of the theory to unusual situations. However,
Griffiths,
and Hartle and Gell-Mann have suggested a more radical use of this decoherence
function, called the consistent histories approach. One of the
 properties of this function is that the probability
of outcome of some individual measurement depends on which other measurements
are assumed to have been made. I.e., the probabilities of the various possible
outcomes of say the operator $B_2$ in the above depends on whether or not
$Z_n$ is included in the history, and on precisely which operator (and
thus physical quantity) $Z_n$ represents. This just corresponds to the
usual quantum condition that the outcome of measurements of say the position
of a particle depends on whether or not I earlier measured the momentum
of that particle, even if I do not know what the outcome of that earlier
measurement was.  In
classical physics of course, the measurement of a quantity need not in
and of itself alter the subsequent behaviour of the system.
The proposal made by  Griffiths, and Gell-Mann and Hartle is that ONLY
those histories which
have the property of `decoherence' (hence the name `decoherence function')
are histories which one can meaningfully talk about. All other histories
are mathematical fictions of the theory. The property of decoherence is
that a complete set of histories $\{{\bf v}\}$ is physically meaningful
only if the decoherence function obeys the property
$$C({\bf v};\tilde{\bf v})= 0$$
if ${\bf v}\neq \tilde{\bf v}$. (Griffiths chooses the weaker requirement
that only the real part of $C$ need obey this condition.) What this condition
means is that
for this set of histories one does not need to worry about whether or not
some attribute has actually been measured, or included in the
history. The probability of the outcomes
of one set of measurements is independent of whether or not one includes
that other measurement in the history. The system behaves classically
in that the measurements
do not affect the subsequent behaviour of the system {\bf as indicated
by the probabilities of the restricted class of measurements included in
that set of histories}.

Let us express this condition formally. Define ${\bf v'}$ in relation to
 {\bf v} by omitting one of the elements of the history.
$${\bf v'}= (a_1,...,p_h,s_j,...,z_n)$$ where
$${\bf v}=(a_1,...,p_h,r_i,s_j,...,z_n)$$
Assume furthermore that $r_i$ is in $Q({\bf v})$, and that we want
 the probability of some specific subhistory  $Q({\bf v'})$
independent of the value of $r_i$. Now given $Prob(Q',r_i)$,
 the probability of $Q'$ independent of the value of $r-i$ is just
\begin{eqnarray}
Prob(q')&=\sum_{r_i}Prob(Q',r_i)\nonumber\\
&= \sum_{r_i} {Tr(C({\bf v})\rho_I C({\bf v}))
\over \sum_Q Tr(C({\bf v})\rho_I C({\bf v}))}\nonumber\\
&= \sum_{r_i} \sum(\tilde r_i) {Tr(C({\bf v})\rho_I C({\bf \tilde v}))
\over \sum_Q Tr(C({\bf v})\rho_I C({\bf v}))}
\nonumber\\
&= {Tr(C({\bf v'})\rho_I C({\bf \tilde v'}))
\over \sum_{Q'} Tr(C({\bf v'})\rho_I C({\bf v'}))}\nonumber
\end{eqnarray}
where $\tilde{\bf  v}$ in the second line is {\bf v} with
 the element $r_i$ replaced by $\tilde r_i$. The second
 line is valid because by assumption $D({\bf v,\bf \tilde v} =0$
 unless ${\bf \tilde v}={\bf v}$. (Although the argument is slightly
 less direct in the Griffiths formulation it is still true there
 as well). The consistent histories formulation can therefore be
 phrased as the statement that
only those histories are physically real for which the
probabilities are independent of whether or not one of the
unknown elements of the history is present or not.

How do the four roles of time enter into this formalism. The first
 two are present in exactly the same way they are in ordinary quantum
theory. The third, the exclusivity and completeness are included by
demanding that the  sum in $C({\bf v})$ of all of the possible values
for  one of the elements of {\bf v} is the same as the $C$ function
excluding that element. I.e., we have
$$\sum_{r_i}C(a_1,...q_h, r_i,s_j,...,z_n)=C(a_1,...q_h,s_j,...,z_n)$$
And that if say $r_i$ and $s_j$ are eigenvalues of the
 same item in the history, at the same time, ($t_i=t_j$), then
$$ C(a_1,...q_h, r_i,s_j,...,z_n)=0$$
unless $ r_i=s_j$. I.e.,, at the same moment in time, the same
item in the history cannot have two different values.

This decoherence function formalism is clearly designed to
implement the fourth condition. Time enters in a crucial way
 because the answers obtained depend crucially of the time
ordering of the projection operators in the definition of $C$.
 If we reverse the order, or any pair, the probabilities will not be the
same.

It is of interest \cite{GMHT} that one can also generalize the
above formula to include the setting of final conditions in a
final density matrix. I.e., we can generalize the formula by
including a $\rho_F$ as a final condition.
$$D'= {Tr(\rho_F C({\bf v}\rho_IC^\dagger({\bf v}
\over(\sum_{Q(\bf v)} Tr(\rho_F C({\bf v}\rho_IC^\dagger({\bf v})}.
$$
Again $\rho_F$ will represent either a prejudice about the
probabilities of the final state of the system (e.g.,
one may want to calculate the probabilities in ones experiment conditional
on the laboratory
still existing at the end of the experiment) or the
accumulated effect of the parts of $K({\bf v})$ which occur
 after all of the times associated with the $Q({\bf v})$.


\begin{references}
\bibitem{curved} It is often said that Einstein's theory describes gravity
by means of a curved spacetime. Note that the model of the beach ball is
 in fact curved. The primary attribute of gravity, that it explains
why things that go up come down again, does not require a curved
spacetime. It is fitting together the various aspects of gravity,
 e.g., that the earth can have a surface which is accelerating
upwards at all points while still retaining the same size, which do
 require a curved spacetime. Furthermore, the curvature is a more
 invariant concept than is the `unequal flow of time' which as I
 mention later has difficulties associated with the definition of
 what one means by time. Unfortunately, the curvature of spacetime
 does not carry very much meaning to the average person, and tends
 to lead to the belief that General Realtivity is impossible to
understand for anyone but some amongst a select few.
\bibitem{Bell-Thm}{} See ref. \cite{Bell}.
See also various articles in ref.\cite{Greenberger},
 especially that by A. Aspect.
\bibitem{know}{} Throughout this paper I use the term `knowledge' instead
of the
more traditional ``measurement" to refer to
the process by which the general formalism
 of the
theory is made to conform to the particulars
 of the world at a specific time, place and condition.
 The latter term I feel causes too much confusion
to be of use in discussions of the foundation
 of quantum mechanics.In the interpretation of quantum mechanics,
 `Measurement'
 has two   separate meanings---
 it can refer to the acquisition of knowledge
about the physical world, or it can refer to
 the physical process which accompanies
that acquisition of knowledge. In the latter case,
 as a physical process, measurements
 are a part of the physical world and should
therefore be analyzable in terms of one's
theory of the physical world ( quantum
mechanics in this case). I will retain the term measurement for this meaning.
 In the former case,
 the term refer to the  activity of making the theory
correspond to the specific peculiarities of
 the physical world one wishes to study, and play a
more axiomatic, and thus unanalyzable, role in the theory.
I will use the term knowledge, instead of measurement, for this meaning.
Thus, in order to avoid the dangers in the confusion between the
physical and
axiomatic roles that the term measurement plays in quantum
theory, I use the term knowledge
throughout for the latter activity and measurement is reserved for the
former meaning. This of course raises its
 own set of problems, as the
controversy about the nature of knowledge is probably
even deeper than controversy about the nature of measurement.
However two seperate terms are needed to differentiate between the two
seperate aspects of the theory usually conflated in the term measurement.
Furthermore, the emphasis that the use of the
term `knowledge' places on the acquisition of information
about the world  as a human activity, is a
more accurate portrayal of the role  played  in quantum theory.
However,
 if the use of the term `knowledge' bothers the reader, it can be
 replaced almost everywhere by the
more traditional term `measurement'.
\end{references}
\end{document}